\documentstyle[12pt]{article}
\newcommand{\eqnum}[1]{\setcounter{equation}{#1}}

\newenvironment{figcaption}[2]{
 \vspace{0.3cm}
 \refstepcounter{figure}
 \label{#1}
 \begin{center}
 \begin{minipage}{#2}
 \begingroup \small Fig. \thefigure: }{
 \endgroup
 \end{minipage}
 \end{center}}

\newcommand{\square}{\kern1pt\vbox{\hrule height
1.2pt\hbox{\vrule width 1.2pt\hskip 3pt
   \vbox{\vskip 6pt}\hskip 3pt\vrule width 0.6pt}\hrule
height 0.6pt}\kern1pt}
\newcommand{\lsim}{\mbox{$\;\lower-.2ex\hbox{$\textstyle<$}
\;\!\!\!\!\!\!
\lower.7ex\hbox{$\textstyle \sim$}\;$}}
\newcommand{\gsim}{\mbox{$\;\lower-.2ex\hbox{$\textstyle>$}
\;\!\!\!\!\!\!
\lower.7ex\hbox{$\textstyle \sim$}\;$}}
\newcommand{\beq}{\begin{equation}}
\newcommand{\eeq}{\end{equation}}
\newcommand{\barr}{\begin{eqnarray}}
\newcommand{\earr}{\end{eqnarray}}
\newcommand{\veck}{\!\!\mbox{ \boldmath $k$}}
\newcommand{\vecx}{\!\!\mbox{ \boldmath $x$}}
\newcommand{\ga}{\mbox{$\gamma$}}
\newcommand{\del}{\partial}

\newcommand{\th}{\mbox{$\theta$}}
\newcommand{\Th}{\mbox{$\vartheta$}}

\newcommand{\Ga}{\mbox{$\Gamma$}}
\newcommand{\Om}{\mbox{$\Omega$}}
\newcommand{\Si}{\mbox{$\Sigma$}}

\newcommand{\si}{\mbox{$\sigma$}}
\newcommand{\vecr}{\!\!\mbox{ \boldmath $r$}}
\newcommand{\vecq}{\!\!\mbox{ \boldmath $q$}}

\begin{document}
\renewcommand{\Large}{\large}
\renewcommand{\huge}{\large}

\begin{titlepage}
\baselineskip .15in
\begin{flushright}
WU-AP/60/96
\end{flushright}

~\\

\vskip 1.5cm
\baselineskip .33in
\begin{center}
{\large\bf  Dynamics of Quiet Universes}

\vskip 0.8cm
\renewcommand{\thefootnote}{\fnsymbol{footnote}} {\sc Hiraku}
MUTOH, {\sc Toshinari} HIRAI\footnote{e-mail: hirai@cfi.waseda.ac.jp} and 
{\sc Kei-ichi} MAEDA\footnote{e-mail: maeda@cfi.waseda.ac.jp}\\
{\em Department of Physics, Waseda University, Shinjuku-ku, Tokyo 169,
Japan}
\renewcommand{\thefootnote}{\arabic{footnote}}
\setcounter{footnote}{0}
\end{center}

\vskip 0.5cm
\vfill
\begin{abstract}
We study the  stability of a contracting {\it silent universe}, which is a
spacetime with irrotational dust and vanishing magnetic part of the Weyl
tensor, $H_{ab}=0$. Two general relativistic backgrounds are analyzed; one
is an attractor of {\it silent universes}, i.e., a locally Kasner
spacetime, and the other is a particular class of inhomogeneous Szekeres
solutions. In both cases their stabilities  against   perturbations with
non-zero magnetic part depend on a  contraction configuration; a
spindle-like collapse is unstable, while a pancake-like collapse is still
stable. We also find that a similar instability exists in  spindle collapse
for a Newtonian case.
 We conclude that $H_{ab}=0$ is not a generic ansatz even in general
relativistic dust collapse.\\ \\
{\em Subject headings}: cosmology: theory --- galaxies: formation ---
relativity --- large-scale structure of universe
\end{abstract}

\vfill
\begin{center}
July, 1996
\end{center}
\vfill

\end{titlepage}

\normalsize
\baselineskip = 18pt

\section{Introduction}
\label{sec:Introduction}\eqnum{0}

\baselineskip .3in
 The recent observations of the large scale structure in the universe show
that quite nonlinear wall-like structures or filament-like structures seem
to exist. Such structures may give us important information about the
structure formation process in the universe. It may be necessary to study
nonlinear dynamics in order to know what kind of structure is formed on
large scales.

In Newtonian gravity, the Zel'dovich approximation (Zel'dovich 1970),
which describes well the evolution of inhomogeneities  even in the weakly
nonlinear regime, predicts a pancake collapse. The analysis of a
homogeneous ellipsoid in an expanding universe also supports such a
pancake collapse (White \& Silk 1979). How about for a very large scale
such that  general relativistic effects become important?

In general relativity, however, the analysis of nonlinear dynamics is
quite difficult  because the Einstein equations are highly nonlinear
partial differential equations. It has been restricted to very simple
cases such as the spherically symmetric Tolman-Bondi model (Tolman 1934;
Bondi 1947).  Recently the nonlinear dynamics of irrotational dust with
vanishing magnetic part of the Weyl tensor ($H_{ab}=0$) has
been examined (Barnes \& Rowlingson 1989; Matarrese et al.
1993; Croudace et al. 1994; Bertschinger \&  Jain 1994,
Bruni et al. 1995a, b). Under such a condition, the system
is described by nonlinear ordinary differential equations,
which makes us able to analyze more generic situations.
Such a system has no degree of freedom of gravitational waves, no
information can transfer between the neighborhoods,  
and each fluid element evolves independently. 
Therefore such 
models were called {\it silent universes} by Matarrese et al. (1994).
Bruni et al. (1995a) found attractor solutions in that system and showed
that a generic gravitational contraction leads to a triaxial spindle-like
configuration.

However the assumption  of $H_{ab}=0$ is controversial. It may
not be necessarily justified for generic 
cases. In fact, Matarrese et al. (1993) and Kojima (1994) 
pointed out, using a perturbative approach in an expanding
Friedmann-Robertson-Walker (FRW) universe, that even if
$H_{ab}=0$ in a linear regime, $H_{ab}$ will be 
generated during the nonlinear evolution of inhomogeneities
and would affect the dynamics of the universe. This 
leads to another question, i.e., ``Is the  attractor in the
silent universe still some attractor in more generic
spacetimes?"

In this paper, we study perturbations of {\it
silent universes}, taking into account the effect of non-vanishing
$H_{ab}$. We shall call such a perturbed spacetime a quiet
universe.  Then the question  above becomes ``Will a quiet
universe be silent ?"  Analyzing stability of the attractor solution
of silent universes
(Kasner-type spacetime) and some particular class of Szekeres
solutions (Szekeres 1975), we discuss whether or not the {\it
silent universe} is generic in general relativity  and what kind of
configuration is more probable in the nonlinear regime of structure
formation.

This paper is organized as follows. In \S 2, we
write down the basic equations for the dynamics of irrotational
dust with the introduction of dimensionless variables and we summarize
the  {\it silent universes}. In \S 3, the stability
of {\it silent universes} against linear
perturbations with non-vanishing $H_{ab}$ is analyzed. We consider
two background spacetimes as unperturbed spacetimes: the Kasner-type
universe and some Szekeres universes and show the dependence of
stability on contraction configuration. We also discuss the 
stability of the Zel'dovich solution in Newtonian
theory in \S 4. Our conclusion follows in \S 5.

We take $c=8\pi G =1$, and the signature  of ($-, +, +, +$).
Throughout this paper, Latin indices are used for 
coordinate components;
$a, b, c, \cdots$ run from 0 to 3 and $i, j, k, \cdots$
 are for the three spatial  
indices, while Greek indices are used for tetrad components;
$\alpha, \beta,
\gamma, \cdots$ run from 0 to 3 and $\lambda, \mu, \nu, 
\cdots$
 are for spatial triad indices. The brackets () and [] denote
symmetrization and anti-symmetrization for all indices.


\section{Dynamics of Irrotational Dust}
\label{sec:Dynamics}
\eqnum{0}

\subsection{Basic Equations for Irrotational Dust}
\label{sec:basiceqs}

The irrotational dust fluid system with four-velocity
$u^{a}$ is described by the following variables: the mass
density $\rho$, the expansion scalar
$\theta$, the shear tensor
$\sigma _{ab}$ and  electric and magnetic parts of the
Weyl tensor $C_{abcd}$; $E_{ab}$ and
$H_{ab}$. Those are defined as
\barr
\theta
&=&u^{c}_{~;c},\\
\sigma _{ab}&=&u_{(a;b)}-{1 \over 3} h_{ab}\theta ,\\
E_{ab}&=&C_{acbd}u^{c}u^{d},\\
H_{ab}&=&^{*}C_{ac
bd}u^{c}u^{d},
\earr
where
$h_{ab} \equiv g_{ab}+u_{a}u_{b}$ is a projection tensor, and 
$^{*}C_{abcd} \equiv \frac{1}{2} \epsilon_{abef}C^{ef}_{~~cd}$
 is the dual of  the
Weyl tensor. 
The basic equations (Ellis
1971) are 
\barr
&&\dot{\theta}+\frac{1}{3}\theta ^2+2\sigma
^2+\frac{1}{2}\rho =0, \\
\label{eqn:RayN}
&&\dot{\rho}+ \theta\rho=0, \\
\label{eqn:conN}
&&\dot{\sigma}_{ab}+\sigma_{ac}\sigma^{c}_{~b}
+\frac{2}{3}\theta
\sigma_{ab} -\frac{2}{3}h_{ab}\sigma ^2+E_{ab}=0,\\
&&h^{ac}h^{bd}\dot{E}_{cd}+h^{c(a} \epsilon
^{b)def} u_{d}H_{ce;f}+h^{ab}\sigma ^{cd}E_{cd}+\theta
E^{ab}-3E_{c}^{~(a}\sigma ^{b)c} \nonumber\\
&&\qquad =-\frac{1}{2} \rho\sigma ^{ab}, \\
&&h^{ac}h^{bd}\dot{H}_{cd}-h^{c(a}\epsilon ^{b)def}u_{d}E_{ce;f} +h^{ab}\sigma
^{cd}H_{cd} +\theta H^{ab}-3H_{c}^{~(a}\sigma ^{b)c}\nonumber\\
&&\qquad =0,
\\  &&\frac{2}{3}\theta^{;a} -\sigma^{ab}_{~~;b}=0, \\
&&H_{ab}=-h_{ac}h_{bd}{\sigma 
^{~(c}_{e}}_{;f}\epsilon^{d)gef}u_{g}, \\ & &
h^{ac}h^{bd}E_{cb;d}-\epsilon ^{abcd}u_{b}
\sigma_{~c}^{e}H_{de}=\frac{1}{3}h^{ab}\rho_{;b}, \\
&&h^{ac}h^{bd}H_{cb;d}+\epsilon ^{abcd}u_{b}\sigma_{~c}^{e}E_{de}=0, 
\earr
where, an overdot represents a directional derivative
along the fluid worldlines, e.g.,
$\dot{T}_{a\cdots b}=T_{a\cdots b;c}u^{c}$.

We introduce the orthonormal tetrad $\{e^{~a}_{(\alpha)}\}$ by 
\beq
e^{~a}_{(0)}=u^{a}=(1, 0, 0, 0),\, \, \,  e^{~a}_{(\mu)}= (0,
e^{~i}_{(\mu)}),
\eeq
with
\beq
e^{~i}_{(\mu)}e^{(\nu)}_{~i}=\delta _{\mu}^{~\nu}.
\eeq
The spatial triad vectors $\{e^{~a}_{(\alpha)}\}$ are chosen to
be  Fermi-propagated along $u ^{a}$, i.e., 
\beq
e^{(\alpha)}_{~a;b}u^{b}=0.
\eeq
The metric is described by those tetrad vectors as,
\barr
ds^2 &= &g_{ab}dx^{a}dx^{b} \nonumber \\
 & =& -dt ^2  +e_{i(\mu)}
e^{~(\mu)}_{j}dx^{i}dx^{j}.
\earr
To write down the basic equations in terms of the tetrad components,
we use the following relation for the covariant derivatives:
\barr
T_{\alpha \beta;\gamma}& = & T_{ab;c}
e^{a}_{~(\alpha)}e^{b}_{~(\beta)}e^{c}_{~(\gamma)} \nonumber \\
             & = & \partial  _{\gamma}T_{\alpha \beta} 
- \Ga^{\delta} _{\ \gamma\alpha} T_{\delta\beta} - 
\Ga^{\delta}_{\ \gamma\beta} T _{\alpha \delta},
\earr
where $\Ga _{\alpha\beta\gamma}\equiv e^{a}_{~(\alpha)}e_{(\gamma) a
; b}e^{b}_{~(\beta)}$ is the Ricci rotation coefficient and $\partial
_{\alpha}$ represents the directional derivative with respect to
${\bf e}_{(\alpha)}$. The governing equations for the tetrad
components  then  become the  same ones for the
coordinate components in Ellis (1971).

We also need the equations for the Ricci rotation coefficients $\Ga ^{\alpha}_{\
\beta\gamma}$ or the structure constants $ 
\ga^{\alpha}_{\ \beta\gamma}=
\Ga^{\alpha}_{\ \beta\gamma} - \Ga^{\alpha}_{\ \gamma\beta}$ to
close our system.
Those equations are obtained from the Jacobi identities, i.e., 
\beq
\partial _{[\delta}\ga ^{\alpha} _{\ \gamma\beta]}+\ga ^{\epsilon}_{\,
[\delta\gamma}\ga ^{\alpha}_{~\beta]\epsilon}=0.
\eeq
From our choice of the tetrad,  some of the components of
$\Ga ^{\alpha}_{~\beta\gamma}$ are known as 
\beq
\Ga ^{0}_{~00}=0, ~~~\Ga ^{\mu}_{~00}=0, ~~~\Ga
^{0}_{~0\nu}=0, ~~~\Ga ^{\mu}_{~0\nu}=0
\eeq
and
\beq
\ga ^{\mu}_{~\nu 0}=\Ga ^{\mu}_{~\nu 0}=\si
_{\mu\nu} +\frac{1}{3}\delta _{\mu\nu}\theta .
\eeq

Next, we define dimensionless variables, $\Om$, $\Si
_{\mu\nu}$, ${\cal E}_{\mu\nu}$, ${\cal H}_{\mu\nu}$, ${\cal
G}_{\mu\nu\rho}$ and ${\cal C}_{\mu\nu\rho}$ as,
\barr
&&\rho = \frac{1}{3}\Om \theta ^2, \ \si _{\mu\nu}=\Si
_{\mu\nu}\theta,   
\ E_{\mu\nu} = {\cal E} _{\mu\nu}\theta
^2,
\ H_{\mu\nu} = {\cal H}_{\mu\nu}\theta ^2, \ \nonumber\\
&&\Ga _{\mu\nu\rho}=
{\cal G}_{\mu\nu\rho}\theta , \ \ga _{\mu\nu\rho} =
{\cal C}_{\mu\nu\rho}\theta ,
\label{eqn:Dless}
\earr
and
\beq
\Si^2
= {1 \over 2} \Si _{\rho\sigma}\Si
^{\rho\sigma} .
\eeq
Using those variables, an attractor solution in silent
universes becomes a fixed point in a ``phase" space.
Now the basic equations for irrotational dust
fluid are
\barr
\theta^{\, '}& = & \frac{1}{6}(2 +\Om +12\Si^2 )\,\theta , \\
\label{eqn:basici}
\Om^{\, '}& = &\frac{1}{3}(1-\Om -12\Si^2 )\,\Om , \\
\Si^{~'}_{\mu\nu}  &= &  \frac{1}{6} (2-\Om -12\Si^2) \,\Si
_{\mu\nu}+\Si _{\mu\rho}\Si
^{\rho}_{~\nu}-\frac{2}{3}\delta_{\mu\nu}\Si^2+{\cal E}_{\mu\nu}, \\
{\cal E}^{~'}_{\mu\nu}& = &\frac{1}{3}(1  -\Om -12\Si^2
)\,{\cal E} _{\mu\nu}+\delta _{\mu\nu}{\cal
E}_{\rho\sigma}\Si
^{\rho\sigma}-\frac32({\cal E}
_{\mu\rho}\Si ^{\rho}_{~\nu}+\Si _{\mu\rho} {\cal E}
^{\rho}_{~\nu})+\frac{1}{6}\Om \Si
_{\mu\nu} \nonumber \\ &&+\frac{1}{2\theta}(\epsilon
_{\nu\rho\sigma}{\cal H}_{\mu}^{~\rho ;\sigma}+\epsilon
_{\mu\rho\sigma}{\cal H}_{\nu}^{~\rho;\sigma})+\frac{1}{\theta ^2
}\theta ^{;\sigma}(\epsilon _{\nu\rho\sigma}{\cal
H}_{\mu}^{~\rho}+\epsilon _{\mu\rho\sigma}{\cal H}_{~\nu}^{\rho}), 
\label{Edot}\\ 
{\cal H}^{~'}_{\mu\nu}& = &\frac{1}{3}(1 -\Om -12\Si^2 
)\,{\cal
H}_{\mu\nu}+\delta _{\mu\nu}{\cal
H}_{\rho\sigma}\Si
^{\rho\sigma}-\frac32({\cal H}
_{\mu\rho}\Si ^{\rho}_{~\nu}+\Si _{\mu\rho}{\cal H}
^{\rho}_{~\nu})   \nonumber \\
&&-\frac{1}{2\theta}(\epsilon _{\nu\rho\sigma}{\cal E}
_{\mu}^{~\rho;\sigma}+\epsilon _{\mu\rho\sigma}{\cal E} _{\nu}^{
~\rho;\sigma})-\frac{1}{\theta ^2 }\theta ^{;\sigma}(\epsilon _{\nu 
\rho\sigma}{\cal E} _{~\mu}^{\rho}+\epsilon
_{\mu\rho\sigma}{\cal E} _{~\nu}^{\rho}),
\earr
\barr 
&&\hspace{-0.5cm}{\theta ^{;\nu} \over \theta }(\Si
_{\mu\nu}-\frac{2}{3} \delta _{\mu\nu})+{\Si
_{\mu\nu}}^{;\nu} =0, \\
&&\hspace{-0.5cm}{\cal H}
_{\mu\nu} = -\frac{\epsilon
_{\mu\rho\sigma}}{2}({1 \over \theta ^2}\theta ^{;\sigma}\Si
_{\nu}^{~\rho}+\frac{1}{\theta }\Si _{\nu}^{~\rho;\sigma})
-\frac{\epsilon_{\nu 
\rho\sigma}}{2}({1 \over \theta ^2}\theta ^{;\sigma}\Si
_{\mu}^{~\rho}+{1 \over \theta}\Si _{\mu}^{~\rho;\sigma}),
\\
&&\hspace{-0.5cm}{\cal E}^{~~;\nu}_{\mu\nu}+{2 \over \theta}\theta
^{;\nu}{\cal E}_{\mu\nu}= \epsilon _{\mu\rho\sigma}\theta\,
\Si ^{\rho\nu}{\cal H}
^{\sigma}_{~\nu}+\frac{1}{9}(\Om _{;\mu} + {2 \over \theta} \theta
_{;\mu}\Om), \\
&&\hspace{-0.5cm}{\cal H}^{~~;\nu}_{\mu\nu}+{2 \over \theta}\theta
^{;\nu}{\cal H}_{\mu\nu}= -\epsilon _{\mu
\rho\sigma}\theta\,\Si^{\rho\nu}{\cal E}^{\sigma}_{~\nu},
\earr
where prime ($'$) denotes the 
derivative with respect to
$\tau \equiv - \int \theta dt$ and $\epsilon ^{\mu\nu\rho}$
is the Levi-Civita  tensor. $\tau \sim \ln$(a proper volume)$^{-1}$
plays the role of time in the contracting phase.
The Jacobi identities are
\beq
\partial _{[\delta}{\cal C} ^{\alpha} _{\ \gamma\beta]}
+\frac{1}{\theta}
 {\cal C} ^{\alpha} _{\ [\delta\gamma}\partial _{\beta]}\theta
+\theta \,{\cal C} ^{\epsilon}_{\ [\delta\gamma}{\cal C}
^{\alpha}_{\ \beta]\epsilon}=0,
\label{eqn:basicf}
\eeq
which consist of 12 independent equations: 9 dynamical and 3
constraint equations.
Hence we have now  26 evolution equations and 17
constraint equations. 

\subsection{Silent Universes}
\label{sec:Silent}

\baselineskip .3in

The {\it silent universe} is proposed by assuming that  (i) the
fluid is  collisionless dust ($p=0$), (ii) it has no
vorticity ($\omega _{ab} =0$), and (iii) the magnetic part
of the Weyl tensor vanishes ($H _{ab} =0$). Such an ansatz seems
to be plausible because gravitational waves may have nothing to do
with the structure formation process\footnote{It turns out, however,
that a contribution of the  magnetic part of the Weyl tensor may
not also be negligible even in the Newtonian limit (Bertinger and
Hamilton 1994).}. 

$\sigma _{ab}$ and $E _{ab}$ are diagonalized under this
condition as shown by Barnes \& Rowlingson (1989). The two
independent components are diagonal: $\sigma
_{11}$, $\sigma _{22}$ and $E _{11}$ , $E _{22}$. The
metric is written as
\beq
ds^2 = -dt^2 +\sum_{i=1}^{3}l^2 _{i}(\vecx ,
t)(dx^{i})^2 ,
\eeq 
and,
\beq
\frac{\dot{l}_{i}}{l_i}= \frac{1}{3}\theta + \si
_{ii} .  ~~~~~~~~~~~
\eeq
where the $l_i$'s represent the scale factors in the
 $i$-direction and $\sigma_{ii}$ is the $i$-$i$ tetrad component  
of the shear tensor.

For the {\it silent universes}, the
evolution equations are  reduced to the following closed
set of ordinary differential equations.
\barr
\theta ^{\, '}  & = & \frac{1}{6}\left(2 + 
\Om  +12\Si^2 \right) \theta , 
\label{theta}\\
\Om ^{\, '}  & = & \frac{1}{3} \left(1-\Om - 12\Si^2  \right)
\Om, 
\\
\Si _{+}^{~'}  & = & \frac{1}{6}\left(2 -\Om -12\Si^2
 \right)\Si_{+} - {1 \over \sqrt{3}} (\Si_{+} ^2 -\Si_{-}^2) 
+{\cal E} _{+},  \\
\Si _{-}^{~'}  & = & \frac{1}{6}\left(2 -\Om - 12\Si^2
 \right) \Si_{-} +{2 \over \sqrt{3}}\Si_{+}\Si_{-}  + {\cal E}
_{-},  \\ {\cal E} _{+}^{~'}  & = &
\frac{1}{3}\left(1-\Om -12\Si^2 \right) {\cal E}_{+} 
+\sqrt{3}(\Si_{+}{\cal
E}_{+} -\Si _{-}{\cal E} _{-})+\frac{1}{6}\Si _{+}\Om,  \\
 {\cal E}
_{-}^{~'}  & = &
\frac{1}{3}\left(1-\Om -12\Si^2 \right){\cal E}_{-}
-\sqrt{3}(\Si_{+}{\cal E} _{-} +\Si _{-}{\cal
E} _{+})+\frac{1}{6}\Si _{-}\Om ,
\earr
where\footnote{This choice of $\pm$
variables is different from the definition by Bruni et al (1995a),
but is the same as that in usual discussion on Bianchi models.}
\barr
\Si _{+} &=& \frac{\sqrt{3}}{2}(\Si _{11}+ \Si _{22}), \, ~~~~~
 \Si _{-} ~=~ \frac{1}{2}(\Si _{11}-\Si _{22}),\, \nonumber \\
{\cal E} _{+}
& = & \frac{\sqrt{3}}{2}({\cal E} _{11}+ {\cal E} _{22}),
\, ~~~~~ {\cal E} _{-} ~=~ \frac{1}{2}({\cal
E} _{11}- {\cal E} _{22}), 
\label{def:+-}
\earr
and
\beq
\Si^2  =  \Si_{+}^2+\Si_{-}^2.
\eeq
In the present case, the 
equations for ${\cal G} _{\alpha\beta\gamma}$,  or ${\cal C}
_{\alpha\beta\gamma}$ are decoupled. From the condition  of $
\dot{H}_{ab}=0$, we also find a new constraint equation.
However, those are guaranteed to be satisfied from the
dynamical and other constraint equations (Lesame et al. 1994).

Because of the absence of spatial derivatives in the
evolution equations, each fluid element evolves without
influence of the neighborhood. Hence it is called 
a {\it silent universe}. It includes not only homogeneous
Bianchi I spacetimes, but also inhomogeneous spacetimes
such as spherically symmetric Tolman-Bondi spacetimes or
Szekeres solutions.

We see the evolution equations for
5 dynamical variables $ \{\Om$, $\Si
_{\pm}$, ${\cal E}_{\pm}\} $ are closed by themselves.
Analyzing them,  Bruni  et al. (1995a)
find attractor
solutions in 5-dimensional ``phase'' space of $ \{\Om$, $\Si _{\pm}$, ${\cal
E}_{\pm}\} $. $\theta$ is obtained from
eq. (\ref{theta}) after solving those 5 variables.

The attractors in  contracting spacetimes ($\theta \, <\, 
0$) locally correspond to a  family of vacuum Kasner spacetimes.
The set of attractors is a circle in a 2-dimensional 
$\{\Si_{+}$, $\Si_{-}\}$ ``phase'' space (Fig. 1).
 Only three
points on the circle of attractors correspond to  a pancake
collapse, i.e., two of the
$l_{i}$ stay finite but the other one vanishes.
The rest of the points
of the attractors correspond to spindle collapse, i.e., two of
$l_{i}$'s vanish but the other one diverges. Bruni {\it et
al.} (1995a) then concluded that contracting regions  generally tend
to be spindle-like configurations in {\it silent universes}.


\section{Quiet Universes: Magnetic Perturbations of Silent
Universes}
\label{sec:Silentper}\eqnum{0}

In order to generalize the analysis by Bruni et al.
(1995a), we consider linear 
perturbations with $H_{ab}$ of a {\it silent universe}.
First we study the case in which the background solution is an 
attractor of contracting {\it silent universes}, i.e., the
Kasner-type spacetimes in \S\ref{sec:Kasner}. Secondly, as
a background solution, we adopt a planar
inhomogeneous Szekeres solution which approaches a particular
attractor solution of silent universes
(\S\ref{sec:Szekeres}). We 
denote  the
background quantities of the silent universe  by a bar
($\bar{\ }$)  and the perturbed ones with tilde ($\tilde{\ }$),
i.e.,

\barr
&&\theta = \bar{\th}(1+\tilde{\Th}), \quad \Om =\bar{\Om}
+\tilde{\Om}, \quad 
\Si_{\mu\nu} = \bar{\Si}_{\mu\nu}
+\tilde{\Si}_{\mu\nu},
\nonumber \\
&&{\cal E}_{\mu\nu} =
\bar{{\cal E}}_{\mu\nu} +\tilde{{\cal
E}}_{\mu\nu},
\quad {\cal H}_{\mu\nu} =
\tilde{\cal H}_{\mu\nu}, \nonumber \\
&&{\cal G}_{\mu\nu\rho} =
\bar{\cal G}_{\mu\nu\rho}
+\tilde{\cal G}_{\mu\nu\rho}, \quad  {\cal C}
_{\mu\nu\rho}=
\bar{\cal C}_{\mu\nu\rho}
+\tilde{\cal C}_{\mu\nu\rho},
\earr
and 
\beq
\bar{\Si}^2= {1 \over 2}\bar{\Si}_{\mu\nu}\bar{\Si}^{\mu\nu} .
\eeq
Note that  $\bar{\cal H}_{\mu\nu}=0$ because the
background spacetime is a  {\it silent universe}.

The background spacetime we analyze here has at least one
homogeneous direction, which we choose to be $x$ (or 1). 
We then analyze only a single plane-wave mode of perturbations
  with a comoving wave number
$\veck =(k,\, 0,\, 0)$. The spatial gradient $\partial _{1}$ is
replaced as 
\beq
\hat{\partial} _{1} \equiv {1 \over \bar{\th}}\partial _{1}
=\frac{ik}{\bar{\th}l_{1}}.
\eeq

In this case, the perturbation equations, which consist of 26 dynamical
 and 17 constraint equations,  are divided into the following 4
groups, where we have
introduced $\{\bar{\Si}_{\pm}\}$, $\{\tilde{\Si}_{\pm}\}$, $\{\bar{\cal
E}_{\pm}\}$, 
$\{\tilde{\cal E}_{\pm}\}$, and $\{\tilde{\cal H}_{\pm}\}$  in the same way
as Eq. (\ref{def:+-}).
\\[.5em] 
{\bf (1) Group 1} ($\tilde{\Th}$, $\tilde{\Om}$, $\tilde{\Si} _{\pm}$, 
$\tilde{{\cal E}} _{\pm}$, $\tilde{{\cal H}} _{23}$, $\tilde{{\cal C}} _{331}$, 
$\tilde{{\cal C}} _{221}
$):\\
\underline{(1-1) 9 dynamical equations}:
\barr
\tilde{\Th} ^{\, '}  & = &  \frac{1}{6}(2+
\bar{\Om}+12\bar{\Si}^2 )\,\tilde{\Th }+\frac{1}{6}\tilde{\Om} 
 + 4 (\bar{\Si} _{+}  \tilde{\Si} _{+} 
+\bar{\Si}_{-}\tilde{\Si}_{-}),  
 \label{eqn:peri}  \\ 
\tilde{\Om} ^{\, '}  & = &  \frac{1}{3} (1
-\bar{\Om} -12 \bar{\Si}^2)\bar{\Om}\,\tilde{\Th}
+\frac{1}{3}(1  -2\bar{\Om}-12\bar{\Si}^2)\,
\tilde{\Om} 
\nonumber \\ & &
 -8\bar{\Om} (\bar{\Si} _{+} 
\tilde{\Si} _{+}+\bar{\Si}_{-}\tilde{\Si}_{-}), \\
\tilde{\Si} _{+}^{~'}  & = & \left[
\frac{1}{6}\bar{\Si}_{+}
\left(2 -\bar{\Om}-12\bar{\Si}^2 \right)
-\frac{1}{\sqrt{3}}(\bar{\Si} _{+}^2 
-\bar{\Si}_{-}^2) +\bar{{\cal E}}
_{+}\right]\tilde{\Th}   
-\frac{1}{6}\bar{\Si} _{+} 
\tilde{\Om} \nonumber\\
&& +\frac{1}{6}(
2-\bar{\Om}-12\bar{\Si}^2) \tilde{\Si}
_{+}  
-4\bar{\Si} _{+}(\bar{\Si} _{+} \tilde{\Si}_{+}
 + \bar{\Si}_{-}\tilde{\Si}_{-})
-{2 \over \sqrt{3}} (\bar{\Si} _{+}
\tilde{\Si}_{+}-\bar{\Si}_{-}\tilde{\Si}_{-})
\nonumber\\ 
&& +\tilde{{\cal E}} _{+} , \\ 
\tilde{\Si} _{-}^{~'}  & = & \left[
\frac{1}{6}\bar{\Si}_{-}
\left(2 -\bar{\Om}-12\bar{\Si}^2 \right)
+{2 \over \sqrt{3}}\bar{\Si}_{+}\bar{\Si}_{-} +\bar{{\cal E}}
_{-}\right]\tilde{\Th}   
-\frac{1}{6}\bar{\Si} _{-} 
\tilde{\Om}\nonumber\\  
&& +\frac{1}{6}(2-\bar{\Om}-12\bar{\Si}^2)
\tilde{\Si}
_{-}
-4 \bar{\Si} _{-}(\bar{\Si}
_{+}\tilde{\Si}_{+}+\bar{\Si}_{-}\tilde{\Si} _{-}) +{2 \over
\sqrt{3}} (\bar{\Si} _{+} \tilde{\Si}
_{-}+\bar{\Si}_{-}\tilde{\Si}_{+} )  \nonumber\\  
&& +\tilde{{\cal E}}_{-}   , \\ 
\tilde{{\cal E}} _{+}^{~'}  & = &  
\left[\frac{1}{3}
\left(1 -\bar{\Om} -12 \bar{\Si}^2
\right)\bar{{\cal
E}}_{+} 
+\sqrt{3}(\bar{\Si}
_{+}\bar{{\cal E}} _{+}-\bar{\Si}
_{-}\bar{{\cal E}} _{-}) +\frac{1}{6}\bar{\Si}
_{+}\bar{\Om}\right]\tilde{\Th}\nonumber \\ & &
+\frac{1}{6}(\bar{\Si} _{+}
-2\bar{{\cal E}} _{+} ) \tilde{\Om}
\nonumber \\ & &
+\frac{1}{3}(1-\bar{\Om}-12\bar{\Si}^2)\tilde{{\cal E}} _{+}
+\sqrt{3}(\bar{\Si} _{+}\tilde{{\cal E}}
_{+}-\bar{\Si}_{-}\tilde{{\cal E}}_{-})
+\frac{1}{6}\bar{\Om}\tilde{\Si} _{+} \nonumber \\ & &
-(8\bar{\Si} _{+} -\sqrt{3}) \bar{{\cal
E}} _{+}\tilde{\Si} _{+}-(8\bar{\Si}_{-}\bar{{\cal E}}_{+}
+\sqrt{3}\bar{{\cal
E}}_{-})\tilde{\Si}_{-}   +\frac{\sqrt{3}}{2}  \hat{\del} _{1}
\tilde{\cal H} _{23}, \\ 
\tilde{{\cal E}} _{-}^{~'}  & = &  
\left[\frac{1}{3}
\left(1 -\bar{\Om} -12 \bar{\Si}^2
\right)\bar{{\cal
E}}_{-} 
-\sqrt{3}(\bar{\Si} _{+}\bar{{\cal E}} _{-}+\bar{\Si} _{-}\bar{{\cal
E}} _{+})
+\frac{1}{6}\bar{\Si}
_{-}\bar{\Om}\right]\tilde{\Th}
\nonumber \\ & &
 +\frac{1}{6}(\bar{\Si}_{-}-2\bar{{\cal
E}}_{-})
\tilde{\Om}
\nonumber \\ & & + \frac{1}{3}(1-\bar{\Om}-12\bar{\Si}^2 
) \tilde{{\cal E}}
_{-}
-\sqrt{3}(\bar{\Si}_{+}\tilde{{\cal
E}}_{-}+\bar{\Si}_{-}\tilde{{\cal E}}_{+})
 +\frac{1}{6}\bar{\Om}\tilde{\Si}
_{-} \nonumber \\ 
& &  -(8\bar{\Si} _{+}
+\sqrt{3}) \bar{{\cal E}} _{-}
\tilde{\Si} _{+}
-(8\bar{\Si}_{-}
\bar{{\cal E}}_{-} +\sqrt{3}\bar{{\cal E}} _{+}) 
\tilde{\Si} _{-}
-\frac{1}{2} \hat{\del} _{1} \tilde{\cal
H} _{23} , \\ 
\tilde{\cal H} _{23}^{~'}  & = &  \left[\frac{1}{3}(
1-\bar{\Om}-12\bar{\Si}^2)+{\sqrt{3}\over 2}(\bar{\Si}
_{+}+\sqrt{3}\bar{\Si} _{-})\right]
\tilde{\cal H} _{23} + (\sqrt{3}\bar{{\cal E}}_{+}-\bar{{\cal
E}}_{-}) \hat{\del} _{1}
 \tilde{\Th}
 \nonumber \\ & &
 +\frac{1}{2}\hat{\del} _{1}
(\sqrt{3}\tilde{{\cal E}} _{+} -\tilde{{\cal E}} _{-} ) 
 -\bar{{\cal E}} _{-}
\tilde{\cal G} _{221}+\frac{1}{2}(\sqrt{3}\bar{{\cal E}} _{+}
+ \bar{{\cal E}} _{-} )  \tilde{\cal G} _{331}  , \\
 \tilde{\cal C} _{331} ^{~'}  & = & 
  -\left[\frac{1}{6}\bar{\Om} +2\bar{\Si}^2 -{1\over
\sqrt{3}}(\bar{\Si} _{+} +\sqrt{3}\bar{\Si} _{-} ) \right]\,
\tilde{\cal C} _{331}   +{2\over \sqrt{3}}\hat{\del} _{1}
\tilde{\Si} _{+} \nonumber \\
&&~~ -
  \frac{1}{3}(1-2\sqrt{3}\bar{\Si} _{+} ) \hat{\del} _{1}
\tilde{\Th} , \\
 \tilde{\cal C} _{221} ^{~'}  & = & 
  -\left[\frac{1}{6}\bar{\Om}   +2\bar{\Si}^2-{1\over
\sqrt{3}}(\bar{\Si} _{+} +\sqrt{3}\bar{\Si} _{-}) \right]
\,\tilde{\cal C} _{221}   -{1\over \sqrt{3}}\hat{\del} _{1}
(\tilde{\Si} _{+}
 -\sqrt{3}\tilde{\Si} _{-} )
\nonumber \\
&&-\left[\frac{1}{3}+{1\over \sqrt{3}}(\bar{\Si} _{+} -
\sqrt{3}\bar{\Si} _{-})\right] 
\hat{\del} _{1}
\tilde{\Th}   ,
\earr
\underline{(1-2) 3 constraint equations}:
\barr
&&   [(\bar{\Si} _{+} + \sqrt{3}\bar{\Si}
_{-})  -\frac{2}{\sqrt{3}}]
\,\hat{\del} _{1} \tilde{\Th}  +\hat{\del} _{1} (\tilde{\Si} _{+} 
+\sqrt{3}\tilde{\Si} _{-} )
+\sqrt{3}(\sqrt{3}\bar{\Si} _{+}
+\bar{\Si} _{-}) \,\tilde{\cal C} _{331}
\nonumber \\
 && ~~~~~+2\sqrt{3}\bar{\Si}_{-}\tilde{\cal C} _{221}   =    0 
\\ 
&& \tilde{\cal H} _{23}  =
 \frac{1}{2}  [ (\sqrt{3}\bar{\Si} _{+}-\bar{\Si}
_{-})\,\hat{\del} _{1} \tilde{\Th}  +\hat{\del} _{1}
(\sqrt{3}\tilde{\Si} _{+} -\tilde{\Si} _{-} )+(\sqrt{3}\bar{\Si}
_{+}+\bar{\Si} _{-})\,\tilde{\cal C} _{331}
\nonumber \\ 
 && ~~~~~-2\bar{\Si}_{-}\tilde{\cal C} _{221}]
, ~~~~~~~~~~~~~~~~~~~~ \\  
&&  \hat{\del} _{1} (\tilde{{\cal E}} _{+} 
+\sqrt{3}\tilde{{\cal E}} _{-})+2\sqrt{3}\bar{{\cal
E}}_{-}\tilde{\cal C} _{221}+\sqrt{3}(\sqrt{3}\bar{{\cal
E}}_{+}+\bar{ {\cal E}}_{-})
\,\tilde{\cal C} _{331} \nonumber \\ 
&& ~~~~~ +2  
  [(\bar{{\cal E}}_{+}
+\sqrt{3}\bar{ {\cal
E}}_{-} ) -\frac{1}{3\sqrt{3}}\bar{\Om}  ] \,\hat{\del} _{1}
\tilde{\Th}  
\nonumber \\ & & ~~~~~ 
= \sqrt{3}(\sqrt{3}\bar{\Si} _{+}-\bar{
\Si}_{-}  )\tilde{\cal H} _{23} +\frac{1}{3\sqrt{3}}  
\hat{\del} _{1} \tilde{\Om}. 
\earr
{\bf (2) Group 2} ($\tilde{\Si} _{23}$, $\tilde{{\cal E}}
_{23}$, $\tilde{{\cal H}} _{\pm}$, $\tilde{{\cal C}} _{123}$, $\tilde{{\cal C}}
_{231}$, $\tilde{{\cal C}}
_{312}$) :\\ 
\underline{(2-1) 7 dynamical equations}:
\barr
\tilde{\Si} _{23}^{~'}  & = &   \frac{1}{6}
[2-\bar{\Om}-12\bar{\Si}^2-2\sqrt{3}(\bar{\Si} _{+}
+\sqrt{3}\bar{\Si}_{-})]
\tilde{\Si} _{23} +\tilde{{\cal E}} _{23}  , \\ 
\tilde{{\cal E}} _{23}^{~'}  & = & \left[{1\over 3}(1-\bar{\Om}
-12\bar{\Si}^2) + {\sqrt{3} \over 2}(\bar{\Si}
_{+}+\sqrt{3}\bar{\Si} _{-})\right]
\tilde{{\cal E}} _{23}
\nonumber \\
 & &  ~~~+\frac{1}{6}[\bar{\Om} +3\sqrt{3}(\bar{{\cal E}} _{+} +
\sqrt{3}\bar{{\cal E}} _{-})  ]\tilde{\Si}
_{23} -\frac{1}{2}\hat{\del} _{1} (\sqrt{3}\tilde{\cal H}
_{+} -\tilde{\cal H} _{-} ) , \\ 
\tilde{\cal H} _{+}^{~'}  & = & 
\frac{1}{3}(1-\bar{\Om}-12\bar{\Si}^2 
) \tilde{\cal H} _{+}
+\sqrt{3}(\bar{\Si}_{+}
\tilde{\cal H}_{+}-\bar{\Si}_{-}
\tilde{\cal H}_{-}) -\frac{\sqrt{3}}{2} \hat{\del} _{1}
\tilde{{\cal E}} _{23}  \nonumber \\ & & 
-\frac{\sqrt{3}}{2}[(\sqrt{3}\bar{{\cal E}} _{+}+ \bar{{\cal E}}
_{-} )\tilde{\cal G} _{123}+(\sqrt{3}\bar{{\cal E}} _{+}-
\bar{{\cal E}} _{-} )\tilde{\cal G} _{312}  ], \\ 
\tilde{\cal H} _{-}^{~'}  & = & 
\frac{1}{3}(1-\bar{\Om}-12\bar{\Si}^2
) \tilde{\cal H} _{-}
-\sqrt{3}(\bar{\Si}_{+}\tilde{\cal H}_{-}+\bar{\Si}_{-}\tilde{\cal
H}_{+}) +\frac{1}{2 }\hat{\del} _{1}
\tilde{{\cal E}} _{23}   \nonumber
\\ & & ~~~-\frac{1}{2}(\sqrt{3}\bar{{\cal E}} _{+}+ \bar{{\cal E}}
_{-} )\tilde{\cal G} _{123} +\frac12(\sqrt{3}\bar{{\cal E}} _{+}-
\bar{{\cal E}} _{-} )\tilde{\cal G} _{312} -2\bar{{\cal E}} _{-} \tilde{\cal G}
_{231},
\\ 
 \tilde{\cal C} _{123} ^{~'}  & = &   -\left[\frac{1}{6}\bar{\Om}  
+2\bar{\Si}^2 +{2\over \sqrt{3}}(\bar{\Si} _{+}+\sqrt{3}\bar{\Si}
_{-}) \right] \,
\tilde{\cal C} _{123}  , \\
 \tilde{\cal C} _{231} ^{~'}  & = &   -\left[\frac{1}{6}\bar{\Om} 
 +2\bar{\Si}^2 +{2\over \sqrt{3}}(\bar{\Si} _{+}-\sqrt{3}\bar{\Si}
_{-} ) \right]
\tilde{\cal C} _{231}  -\hat{\del} _{1} \tilde{\Si} _{23}  , \\
 \tilde{\cal C} _{312} ^{~'}  & = &   -\left(\frac{1}{6}\bar{\Om}
+2\bar{\Si}^2 -{4\over \sqrt{3}}\bar{\Si} _{+} \right)  \tilde{\cal
C} _{312}  + 
\hat{\del} _{1} \tilde{\Si} _{23}  ,
\earr
\underline{(2-2) 4 constraint equations}:
\barr
&  & \tilde{\cal H} _{+}  = -\frac{\sqrt{3}}{2} [   
\hat{\del} _{1} \tilde{\Si} _{23} + \bar{\Si} _{-}\,\tilde{\cal C} _{123}
- \bar{\Si} _{-}\,\tilde{\cal C} _{231} +\sqrt{3}\bar{\Si} _{+}\,\tilde{\cal C}
_{312}], \\  
&  &  \tilde{\cal
H} _{-}  =   \frac{1}{2}  [ \hat{\del} _{1}
\tilde{\Si} _{23}  - (\sqrt{3}\bar{\Si} _{+}+2\bar{\Si}
_{-})\,\tilde{\cal C} _{123}+(\sqrt{3}\bar{\Si} _{+}-2\bar{\Si}
_{-})\,\tilde{\cal C} _{231} 
\nonumber\\&& ~~~~~ 
+\bar{\Si} _{-}\,\tilde{\cal C} _{312} ],
\\  
 &&  \hat{\del} _{1} (\tilde{\cal
H} _{+} +\sqrt{3}\tilde{\cal H} _{-} )  = \sqrt{3} (\sqrt{3}\bar{{\cal E}}
_{+}-\bar{{\cal E}} _{-})   \tilde{\Si} _{23}- \sqrt{3}(\sqrt{3}\bar{\Si}
_{+}-\bar{\Si} _{-})\tilde{{\cal E}} _{23},
\\ 
 &&\hat{\del} _{1}   \tilde{\cal C}_{123}    
=  0
\earr
{\bf (3) Group 3} ($\tilde{\Si} _{31}$, $\tilde{{\cal E}}
_{31}$, $\tilde{{\cal H}} _{12}$, $\tilde{{\cal C}} _{223}$, $\tilde{{\cal C}}
_{113}
$):\\
\underline{(3-1) 5 dynamical equations}:
\barr
\tilde{\Si} _{31}^{~'}  & = &   \frac{1}{6}
[2-\bar{\Om}-12\bar{\Si}^2-2\sqrt{3}(\bar{\Si} _{+}
-\sqrt{3}\bar{\Si}_{-})]
\tilde{\Si} _{31} +\tilde{{\cal E}} _{31},  ~~~~~~~~~~~~~~~ 
\\
\tilde{{\cal E}} _{31}^{~'}  & = &  \left[{1\over 3}(1-\bar{\Om}
-12\bar{\Si}^2) + {\sqrt{3} \over 2}(\bar{\Si}
_{+}-\sqrt{3}\bar{\Si} _{-})\right]
\tilde{{\cal E}} _{31}  \nonumber \\ && ~~~
 +\frac{1}{6}[\bar{\Om}+3\sqrt{3}(\bar{{\cal E}} _{+} -\sqrt{3}
\bar{{\cal E}} _{-})  ] \tilde{\Si} _{31}  
-\frac{1}{2}  \hat{\del} _{1} \tilde{\cal H} _{12},
\\
\tilde{\cal H} _{12}^{~'}  & = &  
\frac{1}{3}(1-\bar{\Om}-12\bar{\Si}^2  -3\sqrt{3}\bar{\Si} _{+}
) \tilde{\cal H} _{12}
-\frac{1}{2}
\hat{\del} _{1}\tilde{{\cal E}} _{31} \nonumber \\ & & ~~~
+\frac{1}{2}[(\sqrt{3}\bar{{\cal E}} _{+}+ \bar{{\cal E}} _{-}
)\tilde{\cal G} _{113}-(\sqrt{3}\bar{{\cal E}} _{+}- \bar{{\cal E}}
_{-} )\tilde{\cal G} _{223} ] , \\ 
\tilde{\cal C} _{223} ^{~'}  & = &   -\left(\frac{1}{6}\bar{\Om} 
 +2\bar{\Si}^2
+{2\over \sqrt{3}}\bar{\Si} _{+} \right)  \,\tilde{\cal C} _{223}  ,
\\
 \tilde{\cal C} _{113} ^{~'}  & = &  
 -\left(\frac{1}{6}\bar{\Om} +2\bar{\Si}^2 + {2\over
\sqrt{3}}\bar{\Si} _{+} \right) 
\,\tilde{\cal C} _{113} + \hat{\del} _{1} \tilde{\Si} _{31} , 
\earr
\underline{(3-2) 5 constraint equations}:
\barr
 &&\hat{\del} _{1}
\tilde{\Si} _{31}-(\sqrt{3}\bar{\Si} _{+}+\bar{\Si} _{-}) 
\,\tilde{\cal C} _{113} -(\sqrt{3}\bar{\Si} _{+}-\bar{\Si}
_{-})\tilde{\cal C} _{223}   =    0, 
\\
&& \tilde{\cal H} _{12}  = - \frac{1}{2} [  
\hat{\del} _{1}
\tilde{\Si} _{31} +(\sqrt{3}\bar{\Si} _{+} -\bar{\Si}
_{-})\,\tilde{\cal C} _{223} -(\sqrt{3}\bar{\Si} _{+}+\bar{\Si}
_{-})\,\tilde{\cal C} _{113} ], ~~~~~~~~~~~~~~~~\\  
  && 
\hat{\del} _{1}
\tilde{{\cal E}} _{31}  -(\sqrt{3}\bar{{\cal E}} _{+}-\bar{{\cal E}}
_{-})\,\tilde{\cal C} _{223} -(\sqrt{3}\bar{{\cal E}}
_{+}+\bar{{\cal E}} _{-})\,\tilde{\cal C} _{113} =   
2\bar{\Si}_{-}\tilde{\cal H}_{12},
\\
 &&   \hat{\del} _{1} \tilde{\cal H} _{12} 
  =    -(\sqrt{3}\bar{{\cal E}}
_{+}+\bar{{\cal E}} _{-}) \tilde{\Si} _{31}  +(\sqrt{3}\bar{\Si}
_{+}+\bar{\Si} _{-})
\tilde{{\cal E}} _{31} , \\ 
&&  \del_{1}  \tilde{\cal C} _{223}
  = 0
\earr
{\bf (4) Group 4} ($\tilde{\Si} _{12}$, $\tilde{{\cal E}}
_{12}$, $\tilde{{\cal H}} _{31}$, $\tilde{{\cal C}} _{112}$, $\tilde{{\cal C}}
_{332}
$):\\
\underline{(4-1) 5 dynamical equations}:
\barr
\tilde{\Si} _{12}^{~'}  & = &  \frac{1}{6}
(2-\bar{\Om}-12\bar{\Si}^2+4\sqrt{3} \bar{\Si}
_{+}) \tilde{\Si} _{12}
+\tilde{{\cal E}} _{12}  ,
\\ 
\tilde{{\cal E}} _{12}^{~'}  
& = &  \frac{1}{3}(1-\bar{\Om} -12\bar{\Si}^2 -3\sqrt{3}\bar{\Si}
_{+}) \tilde{{\cal E}} _{12}
-(\sqrt{3}\bar{{\cal E}} _{+} -\frac{1}{6}\bar{\Om}  ) \tilde{\Si}
_{12} \nonumber \\ &&
 +\frac{1}{2} \hat{\del} _{1} \tilde{\cal H} _{31} ,
\\ 
\tilde{\cal H} _{31}^{~'}  & = &   \left[\frac{1}{3}(
1-\bar{\Om}-12\bar{\Si}^2)+{\sqrt{3} \over 2}(\bar{\Si}
_{+}-\sqrt{3}\bar{\Si} _{-})\right]
\tilde{\cal H} _{31} +\frac{1}{2}\hat{\del} _{1} \tilde{{\cal E}}
_{12}  \nonumber
\\ & & ~~~-\frac{1}{2}(\sqrt{3}\bar{{\cal E}} _{+}- \bar{{\cal E}}
_{-} ) 
\tilde{\cal G} _{332}- \bar{{\cal E}} _{-} \tilde{\cal G}_{112} 
, \\
\tilde{\cal C} _{112} ^{~'}  & = &   
-\left[\frac{1}{6}\bar{\Om} +2\bar{\Si}^2 -{1\over
\sqrt{3}}(\bar{\Si} _{+} - \sqrt{3}\bar{\Si} _{-} ) \right]
\,\tilde{\cal C} _{112}  +  \hat{\del} _{1} \tilde{\Si} _{12} , \\
\tilde{\cal C} _{332} ^{~'}  & = &  
 -\left[\frac{1}{6}\bar{\Om}  +2\bar{\Si}^2-{1\over
\sqrt{3}}(\bar{\Si} _{+} - \sqrt{3}\bar{\Si}_{-}  ) \right] \,
\tilde{\cal C} _{332} ,
\earr
\underline{(4-2) 5 constraint equations}
\barr
&&  \hat{\del} _{1}
\tilde{\Si} _{12} +(\sqrt{3}\bar{\Si} _{+}  -\bar{\Si}
_{-})\,\tilde{\cal C} _{332}-2\bar{\Si} _{-}\tilde{\cal C}
_{112}    =    0, 
\\
&&\tilde{\cal H} _{31}  =
 \frac{1}{2} [  \hat{\del} _{1} \tilde{\Si}
_{12} - 2\bar{\Si} _{-}\tilde{\cal C}_{112} -    
(\sqrt{3}\bar{\Si} _{+}-\bar{\Si} _{-}) \,\tilde{\cal C} _{332}], 
\\
&&  \del
_{1} \tilde{{\cal E}} _{12}+(\sqrt{3}\bar{{\cal E}}_{+}   -\bar{
{\cal E}}_{-})\,\tilde{\cal C} _{332}-2\bar{ {\cal
E}}_{-}\tilde{\cal C}_{112}  =    -(\sqrt{3}\bar{\Si} _{+}+\bar{
\Si}_{-}  )\tilde{\cal H} _{31} ,
~~~~~~~~~~~~~~~~~~~\\
  && 
\hat{\del} _{1} \tilde{\cal H} _{31}   =   2\bar{{\cal E}}
_{-}\tilde{\Si}_{12} -2\bar{\Si}
_{-}\tilde{{\cal E}}_{12},
\\
 && \hat{\del} _{1} \tilde{\cal C} _{332}   =    0.   
\earr

Since each group is decoupled,  6 (in Group 1) and 3 (in Group 2)
dynamical variables remain free at the initial time. 
Although we have fixed our tetrad system in the background
spacetime, we still have  some freedoms for their choice in perturbed
spacetime. The time is chosen as the proper time of a dust particle,
while the spatial coordinates have been left to be free in
perturbations.  As for the physical dynamical degrees of freedom of
perturbations, we must have scalar and tensor modes, which are
coupled except for the FRW spacetime background. Notice that we
do not have a vector mode because of our irrotational dust.  Since our
perturbations are inhomogeneous only in the 1-direction, the tensor modes
are decoupled into two parts; one is coupled to scalar perturbations
(called even parity mode) and the other is decoupled (called odd
parity mode). The perturbations of Group 1 then consist of scalar and
even parity tensor modes, which have 4 dynamical degrees of freedom. 
The other 2 variables are left still to be free because of the freedom
of coordinate transformation in 2- and 3- directions. The perturbations
of Group 2 consist of 2 physically dynamical freedoms (odd parity
tensor modes) and 1 freedom of coordinate transformation in
1-direction.  There is no degree of freedom in Groups 3 and 4, i.e.,
no physical perturbations.   

Initially,  we will specify
$\tilde{\Om}, \Sigma_{+}, 
\tilde{{\cal E}} _{+}$ and $ \tilde{\cal H} _{23}
$ in Group 1 (corresponding to 4 physical modes).
We set $\tilde{{\cal C}} _{331}=\tilde{{\cal C}}
_{221}=0$,
using remaining gauge freedoms,  and determine other variables from the
constraint equations. 
Then we
analyze stability against the perturbations with 4 physically 
independent initial variables, 
$(\tilde{\Om}$, $\tilde{\Sigma}_{+}$, $\tilde{\cal E}_{+}$, $
\tilde{\cal H}_{23}) = (1+i,0,0,0)$, $(0,1+i,0,0)$, $(0,0,1+i,0)$,
 and $(0,0,0,1+i)$. In Group 2, we determine the initial data from 2 physical
variables in the same way as Group 1.

For the first case of our analysis (\S 3.1), the choice of $x$
direction and a single wave mode does not cause a loss of generality,
because
$x$ is not a special direction in the present problem and   any
linear perturbations can be expanded by Fourier series. As for the
second case (\S 3.2), the perturbations in another homogeneous
direction ($y$) will have the same properties as those in the $x$
direction. However, for the perturbations in the direction of
inhomogeneity ($z$), we may not be able to define  invariant
perturbations because the background itself is inhomogeneous. 
 Therefore we have not analyzed it here.


  \subsection{Perturbations of the Attractor Solutions}
  \label{sec:Kasner}

The attractor solutions of {\it silent universes} are
homogeneous vacuum Kasner  type spacetimes, which are
characterized by a constant value of $\eta$, ($0\leq
\eta < 2\pi$), as
\barr
\bar{\Om} & = & 0,~~~ \bar{\Si}^2 ~=  ~{1 \over 3}, \\
\bar{\Si}_{+} & = & {1 \over \sqrt{3}} \cos \eta, ~~~
\bar{\Si}_{-} ~ = ~ {1 \over \sqrt{3}} \sin \eta ,\\
\bar{{\cal E}}_{+} & = & {\sqrt{3} \over 9} (2 \cos \eta -1)(1+
\cos
\eta), \\
 \bar{{\cal E}}_{-} & =& -{\sqrt{3} \over 9} (2 \cos \eta -1) \sin
\eta,
\earr
and
\beq
\bar{\theta} =\frac{\theta_*}{1+\theta_*(t-t_{*})}, 
\eeq
where $\theta_*$ is the  expansion at an initial time $t=t_*$($\tau=\tau_*$).
$\tau$ is given as
\beq
\tau-\tau_*=-\ln[1+\theta_*(t-t_*)] .
\eeq

 We have analyzed the
behaviors of perturbations for various background values ($\Si
_{+}$) and for several 
 scales of the perturbations ($k^{-1}$).

The typical behavior of the perturbations is shown 
in Fig. 2, in which we set  
the initial values of the perturbations 
to 
$(\tilde{\Om}, \tilde{\Sigma}_{+}, \tilde{\cal E}_{+}, 
\tilde{\cal H}_{23}) = (1+i,0,0,0)$.   We also analyze the other 3 independent initial
data in Group 1, and find similar results.

  The evolution of
perturbations is quite different depending on whether the 
background collapse is spindle-like or pancake-like.  Three pancake points are
$(\bar{\Si}_{+},\bar{\Si}_{-}) = ({1
\over 2\sqrt{3}}, \pm {1 \over 2}),$ and $(-{1 \over \sqrt{3}},
0)$.
 In the case
of a spindle-like collapse, the magnetic part of the Weyl tensor
gets larger and larger, and eventually diverges. The ``{\it
silence}'' is broken.   Other variables, e.g., some components of
shear and the electric part of Weyl tensor  also diverge. This
asymptotic behavior is qualitatively independent of the wave number
 $k$ and background parameter $\eta$. On the other hand, in the
case of a pancake-like collapse, the perturbations do not grow for
any wave number. In pancake collapse, we have also analyzed another two
independent perturbation modes, i.e., Group 2 modes and confirmed that it is
stable.

As a result, we conclude that against generic perturbations with 
$H_{ab}$,  a pancake-like attractor is still stable, but
a spindle-like attractor becomes unstable. 


  \subsection{Perturbations of the Szekeres Solutions}
  \label{sec:Szekeres}

\baselineskip .3in

In the above analysis, since the background spacetime is 
a homogeneous  attractor solution, our result might not
be generic.  It is possible that the evolution of the
perturbations in an inhomogeneous background is different
from that in a homogeneous background.  Therefore we consider
the following particular class of Szekeres solutions (Szekeres 1975;
Goode, S. W., \& Wainwright, J., 1982) as an inhomogeneous
background {\it silent universe}:
\beq
ds^2 = -dt^2 +t^{4/3}\left\{dx^2 +dy^2 
+\left[1-t^{-1}\beta (z)\right] ^2 dz^2 \right\} , 
\eeq
where $\beta (z)$ represents an inhomogeneity in   the
$z$-direction. When $\beta (z)$=0, the spacetime
 is just a  flat
Friedmann-Robertson-Walker (FRW) universe. 

The  dimensionless variables defined  
above for this solution are given as 
\barr
\bar{\theta}  & = &\frac{2t-\beta (z)}{t(t-\beta (z))},
\label{eqn:Szth}\\
\bar{\Om}  & = &1-3\bar{\Si}_{+}^2 ,\\
\bar{{\cal E}} _{+} & = & {1\over \sqrt{3}}\bar{\Si}
_{+}(1+\sqrt{3}\bar{\Si}_{+}) , \\
\bar{{\cal E}} _{-} & = & \bar{\Si} _{-}  ~=~0 ,
\label{eqn:SzSi-}
\earr
where 
\beq
\bar{\Si} _{+}  = \frac{\beta (z)}{\sqrt{3}\left[\beta (z)
-2t\right] }. \label{eqn:SzSi+}
\eeq
$\tau$ is given as
\beq
\tau-\tau_*=-\ln \frac{t(t-\beta (z))}{t_*(t_*-\beta (z))} .
\eeq

There is no  stationary point in the
``phase'' space (see Fig.1). In the case of contracting phase, 
when $t$ decreases, even if the Szekeres universe is 
located initially near the isotropic FRW point  ($\bar{\Si} _{+}=
0$), such spacetimes fall into  different  attractors according
to the sign of
$\bar{\Si} _{+}$, i.e., 
when $-1/\sqrt{3}<\bar{\Si} _{+}<0$, it falls into a pancake-like
attractor ($\bar{\Si} _{+}=-1/\sqrt{3}$), while if 
$0<\bar{\Si} _{+}< 1/\sqrt{3}$, then  it approaches a
spindle-like attractor ($\bar{\Si} _{+}= 1/\sqrt{3}$)
 (see Fig.1). 

Since in the $z$ direction, 
the background Szekeres solution is inhomogeneous, we focus only on
the perturbations in the $x$, $y$ directions.
Here we analyze only a single plane-wave mode  with a wave number
$\veck =(k,\, 0,\, 0)$, as mentioned before.

The results are shown in Figs. 3, in which we set  
the initial values of the perturbations  
$(\tilde{\Om}, \tilde{\Sigma}_{+}, \tilde{\cal E}_{+}, 
\tilde{\cal H}_{23}) = (1+i,0,0,0)$.   We also analyze the other 3
independent initial data in Group 1, and find similar results.
For a spindle-like  background  ($\bar{\Si}_{+} > 0$), the
perturbation of the magnetic part of the Weyl tensor grows
infinitely and the ``{\it silence}'' is broken. The electric part
and the shear also diverge. On the other hand, for a pancake-like
background, the perturbations do not grow. The result depends only
on the contraction configuration, i.e., the sign of
$\bar{\Si}_{+}$. It  does not depend on
 other factors such as the spatial scale of
perturbations. In pancake collapse, we have again analyzed another two
independent perturbation modes, i.e., Group 2 modes and  confirmed that it
is stable.

Similarly to the Kasner-type
background case, the present analysis shows that a pancake-like
collapse is stable while a spindle-like collapse is
unstable. One can also find that an isotropic collapse (FRW) is 
unstable against perturbations with magnetic part $H
_{ab}$ (what is called the tensor mode).

Then some  questions may arise, i.e., ``Where will the 
destabilized  solution  approach in the ``phase space''
after a nonlinear evolution ?" or ``How will the configuration
change?"  Our analysis shows that  the
perturbations with  any wave length eventually approach  the
homogeneous one ($k=0$). Therefore, we may expect that  the
spacetime always evolves  locally into some Bianchi IX spacetime
with complex oscillations near  the singularity as shown in
Belinskii et al (1970, 1982).


 \section{Newtonian Case: Stability of Zel'dovich solutions}
  \label{sec:Zeldovich}\eqnum{0}

Finally, in order to understand our results better, 
we  analyze the stability of  a {\it silent  universe} in 
Newtonian gravity.
The one-dimensional Zel'dovich solution in
Newtonian  gravity is  very similar to the Szekeres solution in
general relativity as pointed out by Kasai (1993), and then
it may be
regarded as a {\it silent universe}.
Adopting
the Zel'dovich solution as a background solution, we  analyze
its perturbations.

In Newtonian gravity, we can describe the irrotational 
dust fluid dynamics  similar to the case in general
relativity except for the absence of the Maxwell-like equations
of $E_{ij}$ and $H_{ij}$ in general relativity (Ellis 1971). The
system with irrotational  dust fluid is described by a velocity
field $v^i$,  mass density $\rho$,
expansion scalar $\theta =v^i_{~,i}$, 
shear tensor $$\sigma
_{ij}=v_{(i,j)}
-{1 \over 3}\delta_{ij}\theta ,$$ and  tidal force
$$E_{ij}= \phi _{,ij}-{1 \over 3}\delta
_{ij}\phi ^{,l}_{~,l},$$
where $\phi$ is Newtonian gravitational potential. 

The
evolution equations of $\rho $, $\theta $, or $\sigma
_{ij}$ have exactly the same expressions in general
relativity, but $E_{ij}$ is determined  only by its
constraint equations without its evolution equation.
$H_{ij}$ does not appear. The equations
for dimensionless variables, $\Om$, $\Si _{ij}$,
${\cal E} _{ij}$, defined by the same  equations
as (\ref{eqn:Dless}), are
\barr
\theta ^{\, '}& = & \frac{1}{6}(2 +\Om +12\Si^2 ) \theta, \\
\Om ^{\, '}& = &\frac{1}{3}(1-\Om-12\Si^2) \Om, \\
\Si ^{~'}_{ij}  &= &\frac{1}{6}(2 -\Om -12\Si^2)
\Si _{ij}+\Si _{il}\Si
^{l}_{~j}-\frac{2}{3}\delta _{ij}\Si
^{2}+{\cal E}
_{ij},
\earr
\barr
&&{\theta_{,j} \over \theta }(\Si ^{ij}
-\frac{2}{3} \delta ^{ij})+\Si_{~~,j}
^{ij} =0, \\
 &&\epsilon^{ilm}(\Si
_{jl,m} + {1 \over \theta}\theta_{,m}
\Si _{jl})+\epsilon _{jlm}(\Si
^{il,m}+{1 \over \theta } \theta
^{,m}\Si ^{il})=0, ~~~~~~~~~~~~~~\\
&&{\cal E}
^{ij}_{~~,j}+{2 \over \theta}\theta
_{,j}{\cal E}^{ij}=\frac{1}{9}(\Om ^{,i}+{2 \over \theta}  \theta
^{,i}\Om), \\
&&\epsilon ^{ilm}({\cal E}
_{jl,m}+{2 \over\theta}\theta
_{,m}{\cal E} _{jl})+\epsilon _{jlm}({\cal E} ^{il,m}+{2
\over\theta} \theta ^{,m}{\cal E} ^{il})=0, 
\earr
where $\Si^2 = \Si_{lm}\Si^{lm}/2$ .

The one-dimensional Zel'dovich solution is  obtained by
transforming  from the Eulerian coordinates $\vecr = ( r_1 , r_2
, r_3 )$ to  the Lagrangian coordinates
$\vecq = ( q_1 , q_2 , q_3 )$;
\barr
r _1 &=& t^{2/3}q_{1} , \\
r _2& =&t^{2/3}q_{2} , \\
r _3 &=& t^{2/3} [q_3 - D(t)S (q_3) ],
\earr
where $S(q_3)$ is an arbitrary function of $q_3 $. Since we are
interested in the contraction phase,  we choose
$D(t)=t^{-1}$ as  $t$ decreases.  This
choice is different from the usual
Zel'dovich solution, in which the growing mode of {\it
linear perturbation theory} in an expanding universe is
chosen ($D(t)=t^{2/3}$ in the case of Einstein-de Sitter 
model). Although the perturbation of the Zel'dovich solution was
 already studied by Bildhauer et al. (1992), we reanalyze it here
because  our choice of $D(t)$ is different from theirs.

Transmission of
information will take place due to the Poisson equation, so 
non-locality exists in general as is well known in Newtonian
gravity. However, the Zel'dovich solution can be called a {\it
silent universe}, because once an initial condition is set,
that is,
$S(q_3)$ is given, each fluid element evolves independently of
other neighbors. 

Therefore, the perturbation analysis can be done in the same way
as  in
\S \ref{sec:Szekeres}. In fact the background variables are given by the
same solutions as eqs. (\ref{eqn:Szth})$\sim$(\ref{eqn:SzSi+}) with
replacement of
$\beta(z)$ with
$S(q_3)$. We
analyze  only perturbations in the 
$x$ direction with a wave number
$\veck =(k,\, 0,\, 0)$.
 The
perturbation equations are
\barr
\tilde{\Th}^{\, '} & = &
{ 1\over 6}\left(2 + \bar{\Omega} +
12{\bar{\Si}_{+}^2} \right) \, 
\tilde{\Th} + 
{1\over 6} \tilde{\Omega}+ 4\, \bar{\Si}_{+}\, \tilde
{\Si}_{+} ,  \\
 \tilde{\Omega}^{\, '} & = &
 \, {1\over 3}\left(1 -  \bar{\Omega} -  12\,\bar{\Si}_{+} ^2 \right)
\bar{\Omega}\, \tilde{\Th}+{1\over 3}\left( 1
-2\bar{\Omega}-12\,{\bar{\Si}_{+}}^2
\right) 
 \tilde{\Omega} \nonumber \\ &&
 -  8\, \bar{\Omega}\, \bar{\Si}_{+}\,
\tilde{\Si}_{+} , 
\\
\tilde{\Si}_{+}^{~'} & = & \left[  \bar{\cal E}_{+} + 
{1\over 6}\left(2 - \bar{\Omega} - 2\sqrt{3} ( 1 +
2\sqrt{3}\,\bar{\Si}_{+})
\bar{\Si}_{+}  
\right)\,\bar{\Si}_{+}  \   \right] \,
\tilde{\Th}
 \nonumber  \\
& &  -   {1\over 6} \bar{\Si}_{+}\,\tilde{\Omega} +\tilde{{\cal E}}_{+} 
+  { 1\over 6}\left[ 2
-  \bar{\Omega} -  
4\sqrt{3}\bar{\Si}_{+}(1+3\sqrt{3}\bar{\Si}_{+}) \right]
\,\tilde{\Si}_{+}  , \\ 
\tilde{\Si}_{-}^{~'} & = & \tilde{{\cal E}}_{-} +{1\over 6}\left[
2 - \bar{\Omega} + 4\sqrt{3}\bar{\Si}_{+}(1 -
\sqrt{3}{\bar{\Si}_{+})} 
\right]  \tilde{\Si} _{-},
\earr
\barr
&&(\sqrt{3}\bar{\Si}_{+}-2) \, 
\tilde{\Th} _{,1} + \sqrt{3}(\tilde{\Si}_{+}+
\sqrt{3}\tilde{\Si}_{-}) _{,1} =  0,   \\
&&3\sqrt{3} (\tilde{\cal E}_{+} +\sqrt{3}\tilde{\cal E}_{-})_{,1}
+2(3\sqrt{3}\bar{\cal
E}_{+}-\bar{\Omega}) \,\tilde{\Th} _{,1} = 
\tilde{\Omega}_{,1}, 
~~~~~~~~~~~~~~~~~~~~~~~~~~~~~~~~~\\ 
&&\sqrt{3}\bar{\Si}_{+}\, \tilde{\Th} _{,1}+
(\sqrt{3}\tilde{\Si}_{+}-\tilde{\Si}_{-}) _{,1} =  0,   \\
&&2\sqrt{3}\bar{\cal E}_{+}\, \tilde{\Th}
_{,1}+(\sqrt{3}\tilde{{\cal E}}_{+}-\tilde{{\cal E}}_{-}) _{,1} = 
0,  \\
&& \tilde{\Si} _{12, 1}=\tilde{\Si} _{13, 1}= \tilde{\Si}
_{23, 1} =\tilde{{\cal E}} _{12, 1}=\tilde{{\cal E}} _{13,
1}=\tilde{{\cal E}} _{23, 1}=0. 
\earr

We have 6 variables with 4 constraint equations.  Then the
degree of  freedom is two, which correspond to scalar 
perturbation modes.  We find that the perturbations of  
$\tilde{\Si} _{ij}$ and $\tilde{\cal E}
_{ij}$ behave very similarly to those in the case of
the Szekeres background in general relativity. For a
spindle-like collapse, the perturbations diverge, while
for a pancake-like background, the perturbations decay
(see Fig. 4, in which we set  
the initial values of the perturbations  
$(\tilde{\Om},
\tilde{{\cal E}} _{+}) = (1+i,0) $.   We
also analyze another set of independent initial data $(\tilde{\Om},
\tilde{{\cal E}} _{+}) = (0, 1+i) $,
and find similar results). This result is consistent with the
preference of pancake collapse  in Newtonian theory.
 The results
suggest that the instability of a spindle collapse shown
in \S
\ref{sec:Silentper} is not due to an extreme situation 
near a singularity in general relativity.


\section{Conclusion}
\label{sec:conclusion}
We have studied the dynamics of a quiet universe, i.e., the
perturbations of a  {\it silent universe} with non-zero 
$H_{ab}$. The analysis of the {\it silent universe} by
Bruni et al. (1995a) shows that a gravitational collapse prefers
a  spindle-like configuration. However, to consider a more
generic case, we have
analyzed  perturbations with non-vanishing $H_{ab}$ for two
general relativistic background solutions: (i) the attractors of
the {\it silent universe}  and (ii) a particular class of 
Szekeres solutions. In both cases, the behaviors of
perturbations are qualitatively the same. The stability clearly
depends on the configuration of a contracting background  
spacetime, i.e., either a pancake-like or a spindle-like
collapse. In the case of a spindle-like collapse, the
perturbations diverge and the {\it silence} is broken. The
spacetime locally approaches the most generic homogeneous
spacetime, that is, the  Bianchi IX universe. On the other hand,
the perturbations do not grow in the case of a pancake-like
collapse.  We conclude that even in general relativity, taking
into account the effect  of
$H_{ab}$, a spindle  collapse is destabilized. 

This result agrees
with the fact that a Newtonian collapse prefers a
pancake-like configuration. Our Newtonian analysis of the
stability of the Zel'dovich solution also supports this result. 
To know the relation between general relativistic perturbations
with non-vanishing magnetic part and the Newtonian ones, we
study the behavior of the magnetic part in the relativistic equation 
(\ref{Edot}) (see Hui 1995).  We find that the terms with the magnetic part 
diverge for a spindle collapse but keep finite for a pancake collapse. 
 This
suggests that the instability for a spindle collapse in Newtonian theory
may also be caused by the magnetic part which is responsible for
transmission of gravitational information. We  conclude that an
ansatz of silence ($H_{ab}=0$) is no longer valid in generic
gravitational contraction.

\vskip 1cm
We would like to thank P. Haines and Y. Kojima for useful
discussions.  This work was supported
partially by the Grant-in-Aid for Scientific Research  Fund
of the Ministry of Education, Science and Culture  (No.
06302021 and No. 06640412),   and by the Waseda University
Grant for Special Research Projects. 

\newpage
\noindent {\large \bf References}
\baselineskip 18pt
\begin{description}
\item Barnes, A., \& Rowlingson, R. R. 1989, Class. Quantum  Gravity, 
{\bf 6}, 949 
\item Belinskii, V. A., Khalatonikov, I. M., \& Lifshitz, E. M. 1970, Adv. 
Phys., {\bf 19}, 525 
\item ------. 1982,Adv.  Phys., {\bf 31}, 639
\item Bertschinger, E., \&  Jain, B. 1994, ApJ, {\bf 431}, 486
\item Bertschinger, E. \&  Hamilton, A. J. S. 1994, ApJ,  {\bf 435}, 1
\item Bildhauer, S., Aso, O., Kasai, M., \& Futamase,T. 1991.  Mon. Not. R.
astr. Soc. {\bf 252}, 132 
\item Bondi, H. 1947, MNRAS, {\bf 107}, 410
\item Bruni, M., Matarrese, S., \& Pantano, O. 1995a, ApJ,  {\bf 445}, 958
\item ------. 1995b, Phys. Rev. Lett.,  {\bf74}, 1916
\item Croudace, K. M., Parry, J., Salopek, D. S., \& Stewart, J.  M. 1994,
ApJ, {\bf 431}, 22 
\item Ellis, G. F. R. 1971, in General Relativity and Cosmology, ed. 
Sachs, R. K. (Academic, London), 104
\item Goode, S. W., \& Wainwright, J.  1982, Phys. Rev. D,  {\bf 26}, 3315 
\item Hui, L., \&  Bertschinger, E. 1995, ApJ, submitted 
\item Kojima, Y. 1994, Phys. Rev. D,  {\bf 50}, 6110
\item Lesame, W. M., Dunsby, P. K. S., \& Ellis, G. F. R. 1995,  Phys.
Rev. D., {\bf 52}, 3406 
\item Matarrese, S., Pantano, O., \& Saez, D. 1993, Phys. Rev. D,  {\bf
47}, 1311 
\item Matarrese, S., Pantano, O., \& Saez, D. 1994, Phys. Rev.  Lett.,
{\bf 72}, 320 
\item Szekeres, P. 1975, Comm. Math. Phys., {\bf 41}, 55
\item Tolman, R. C., 1934, Proc. Natl. Acad. Sci., {\bf 20}, 169
\item White, S. D. M.,  \& Silk, J. 1979, ApJ, {\bf 231}, 1
\item Zel'dovich, Ya. B. 1970, A \& A, {\bf 5}, 84
\end{description}

\newpage
\noindent
{\large \bf Figure Captions}\\
\begin{figcaption}{fig1}{16cm}A set of attractors of  {\it silent
universes}  in the \{$\Si _{+}$,
$\Si _{-}$\} plane; three dots represent  pancake-like attractors. The
rest points on the circle correspond to spindle-like attractors. The line
of
$\Si _{-}=0$ ($-1/\sqrt{3} <
\Si _{+} < 1/\sqrt{3}$) shows a particular class of Szekeres solutions
adopted  as the background solution in \S
\ref{sec:Szekeres}.
  \end{figcaption}

\begin{figcaption}{fig2}{16cm}
The evolution of perturbations for the
attractor  of {\it silent  universes}. The initial values of  the
perturbations are chosen as 
$(\tilde{\Om}, \tilde{\Sigma}_{+}, \tilde{\cal E}_{+}, 
\tilde{\cal H}_{23}) = (1+i,0,0,0)$. We set $\tilde{{\cal C}}
_{221}=\tilde{{\cal C}} _{331}=0$ and other initial values are fixed by
the constraint equations. Figs. 2(a) and 2(b) are in the case of pancake
background ($\bar{\Si}_+=-1/\sqrt{3}$). Figs 2(c), 2(d) and 2(e), 2(f) are
in the case of spindle background ($\bar{\Si}_+=-1/(4\sqrt{3})$ and
$1/\sqrt{3}$, respectively). The solid  and dashed lines show the cases of
the wave number
$k/l_{1*}
\th_* = 10$ and $ =1$, respectively. 
\end{figcaption}

\begin{figcaption}{fig3}{16cm}
Fig. 3: The evolution of perturbations for a
particular class of Szekeres  solutions.  The initial values of the
perturbations are chosen as 
$(\tilde{\Om}, \tilde{\Sigma}_{+}, \tilde{\cal E}_{+}, 
\tilde{\cal H}_{23}) = (1+i,0,0,0)$. We set $\tilde{{\cal C}}
_{221}=\tilde{{\cal C}} _{331}=0$ and other initial values are fixed by
the constraint equations.   Figs. 3(a) and 3(b) are in the case of pancake
background ($\bar{\Si}_+=-1/(4\sqrt{3})$). Figs 3(c) and 3(d)  are in the
case of spindle background ($\bar{\Si}_+=1/(4\sqrt{3})$). The solid  and
dashed lines show the cases of the wave number
$k/l_{1*}
\th_* = 10$ and $ =1$, respectively.
Figs. 3(e) and 3(f) are in the case of isotropic background
($\bar{\Si}_+=0$).  As for the isotropic(FRW) background, as was well
known,  the scalar perturbations are decoupled from the tensor modes.
Hence, although our perturbation with the initial value
$(\tilde{\Om}, \tilde{\Sigma}_{+}, \tilde{\cal E}_{+}, 
\tilde{\cal H}_{23}) = (1+i,0,0,0)$ diverges and the FRW spacetime is
unstable against this perturbation(Fig. 3(e)),
$\tilde{{\cal H}}_{23}$ remains zero because it is a scalar mode. As for 
initial perturbations  with $\tilde{{\cal H}}_{23} \neq 0$, $\tilde{{\cal
H}}_{23}$ diverges and then the FRW spacetime is unstable against this
tensor perturbation as well(Fig. 3(f)). Fig. 3(e) has just a solid line
because the perturbation equations for the scalar mode of isotropic dust
universe do not involve the wave number.
\end{figcaption}

\begin{figcaption}{fig4}{16cm}
The evolution of perturbations for the
Zel'dovich solutions.  The initial values of the perturbations are chosen
as 
$(\tilde{\Om}, \tilde{{\cal E}} _{+}) = (1+i,0) $. Other initial values
are fixed by the constraint equations.   Fig. 4(a)  is in the case of a
pancake background ($\bar{\Si}_+=-1/(4\sqrt{3})$), while Fig. 4(b) is in
the case of a spindle background ($\bar{\Si}_+=1/(4\sqrt{3})$). Figs. 4
are drawn with just a solid line because the perturbation equations do not
involve the wave number.
\end{figcaption}
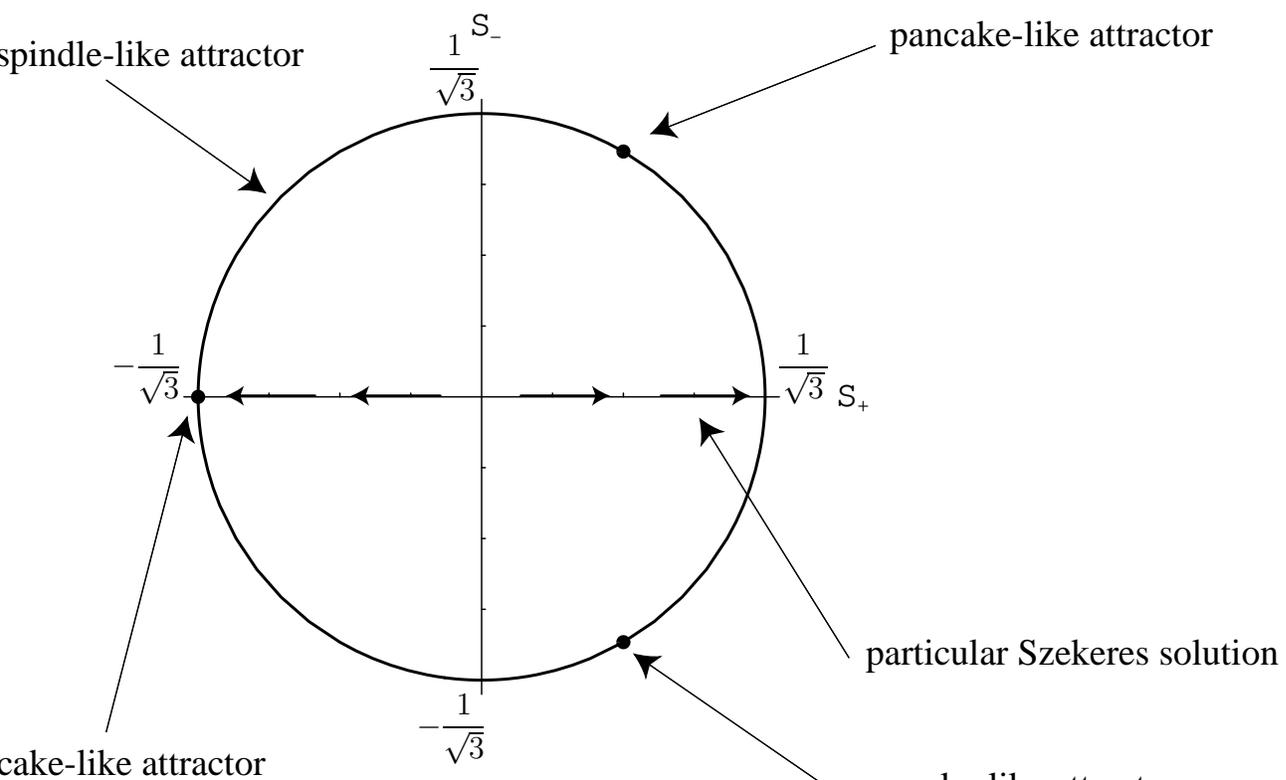
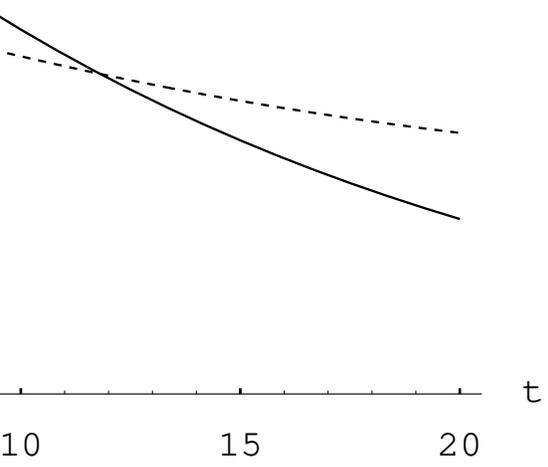
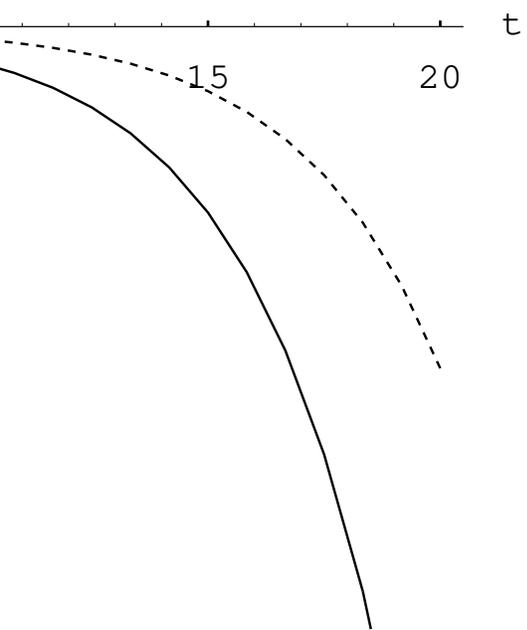
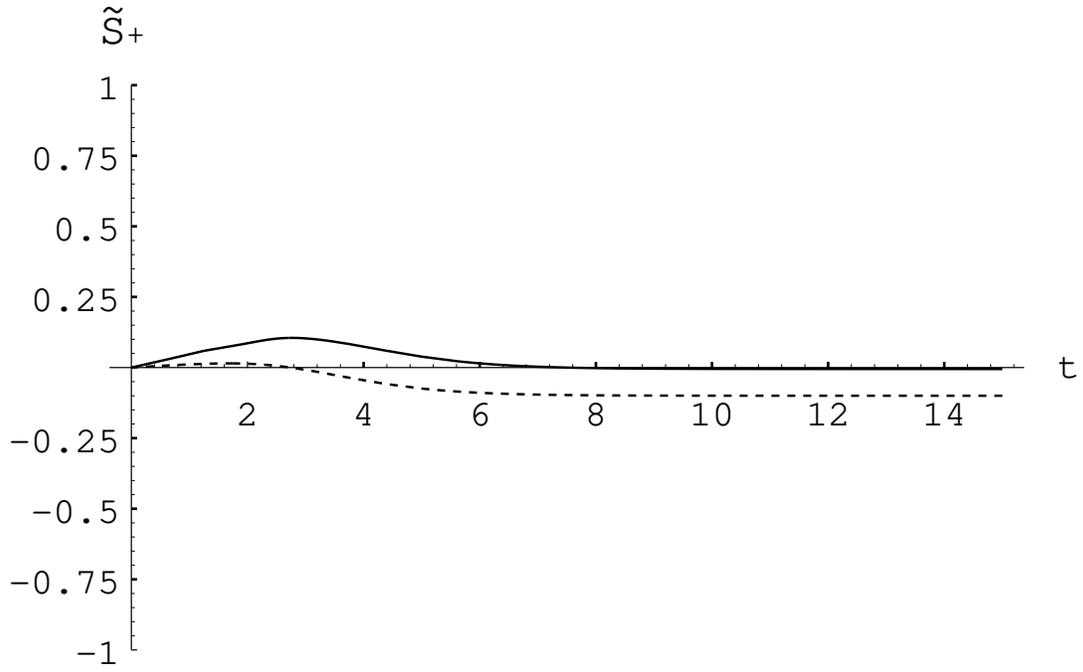
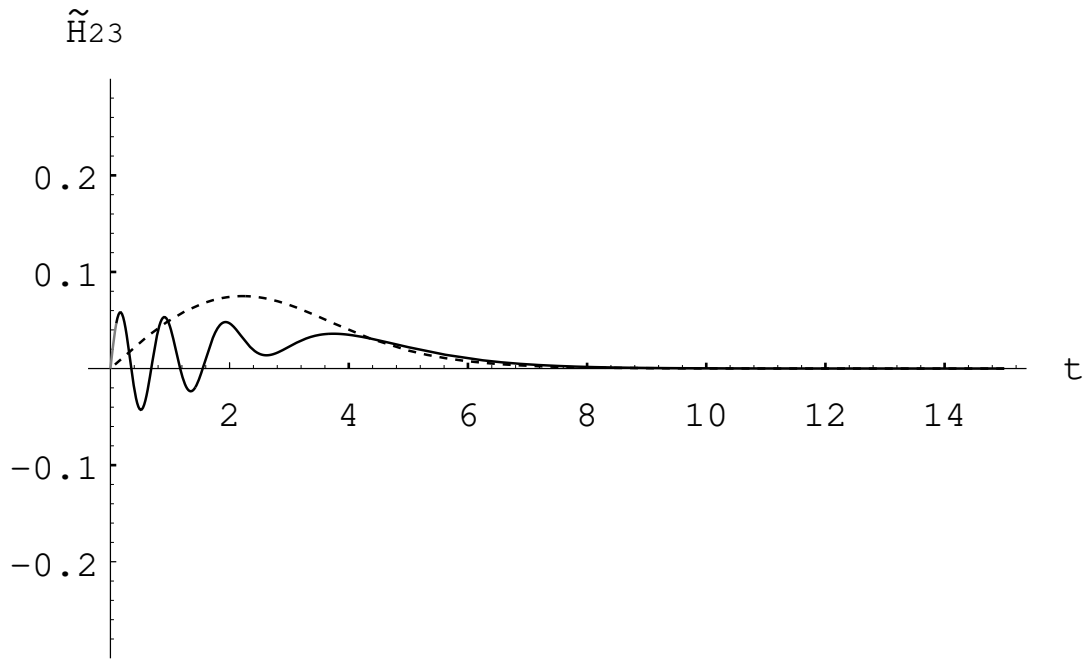
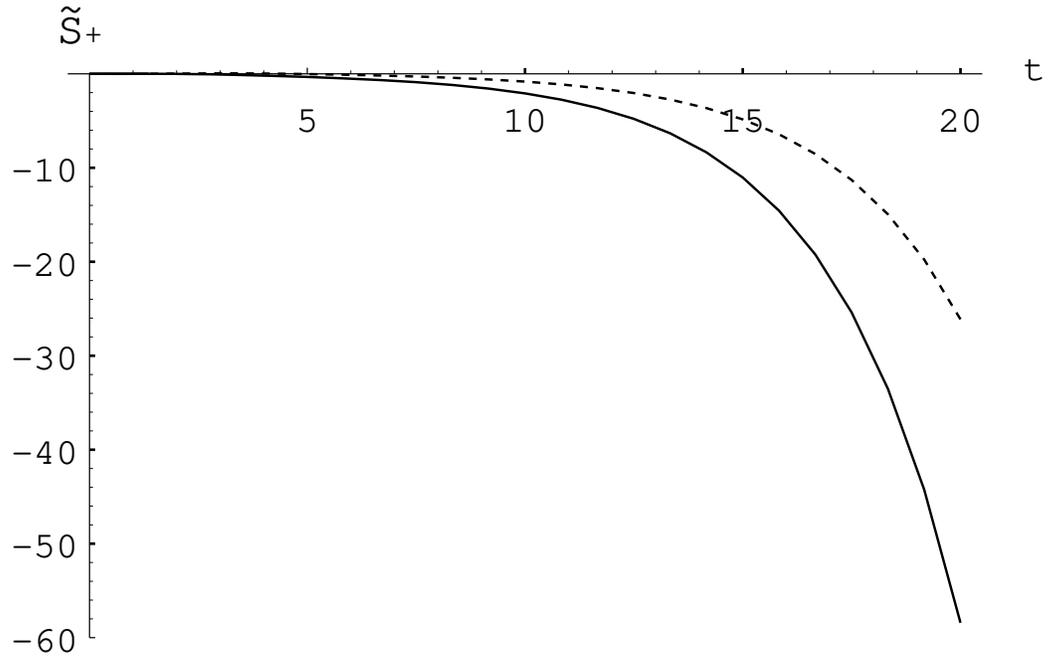
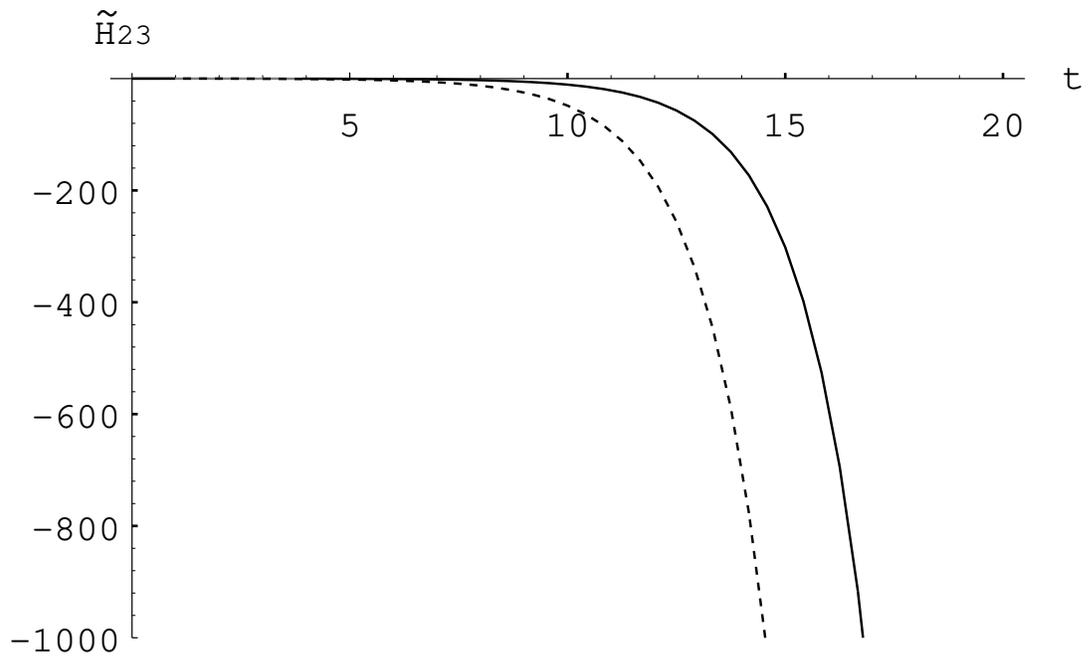
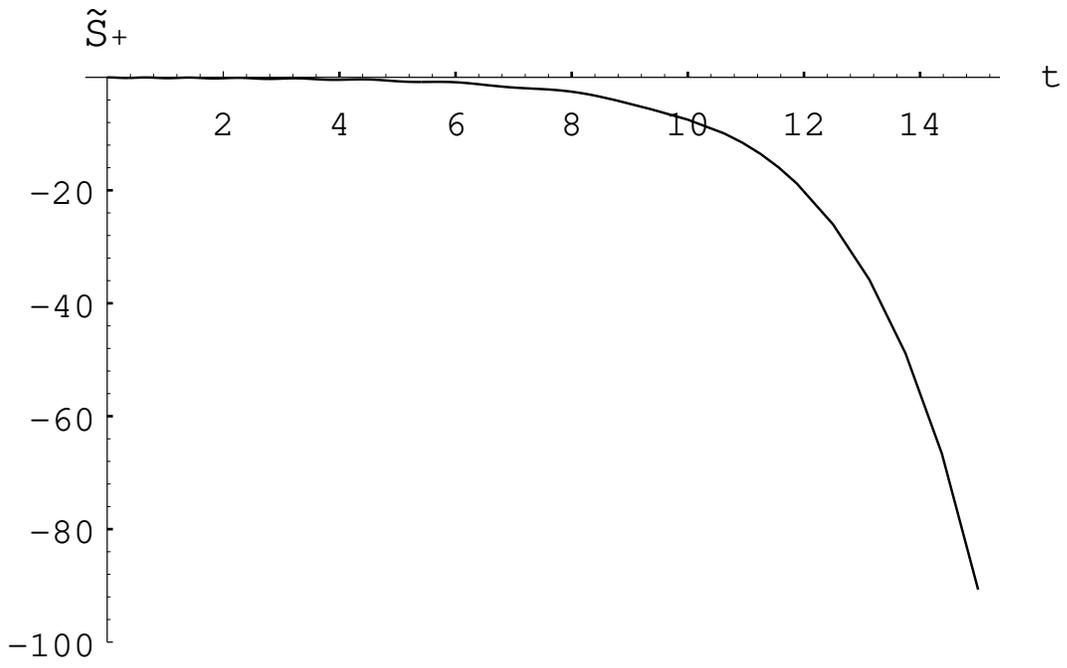
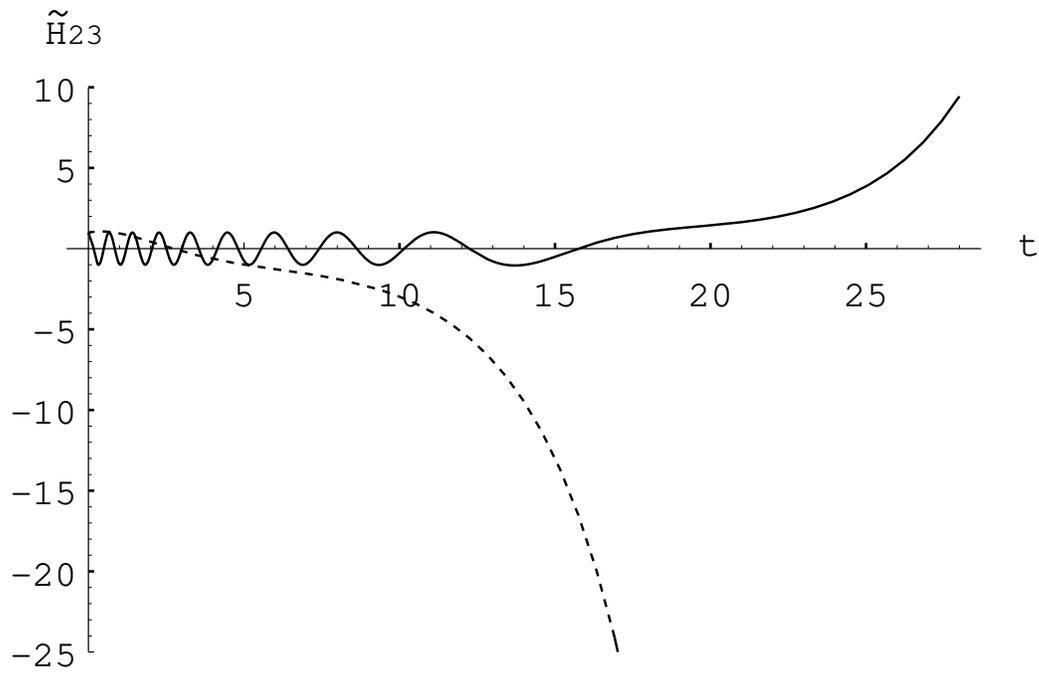
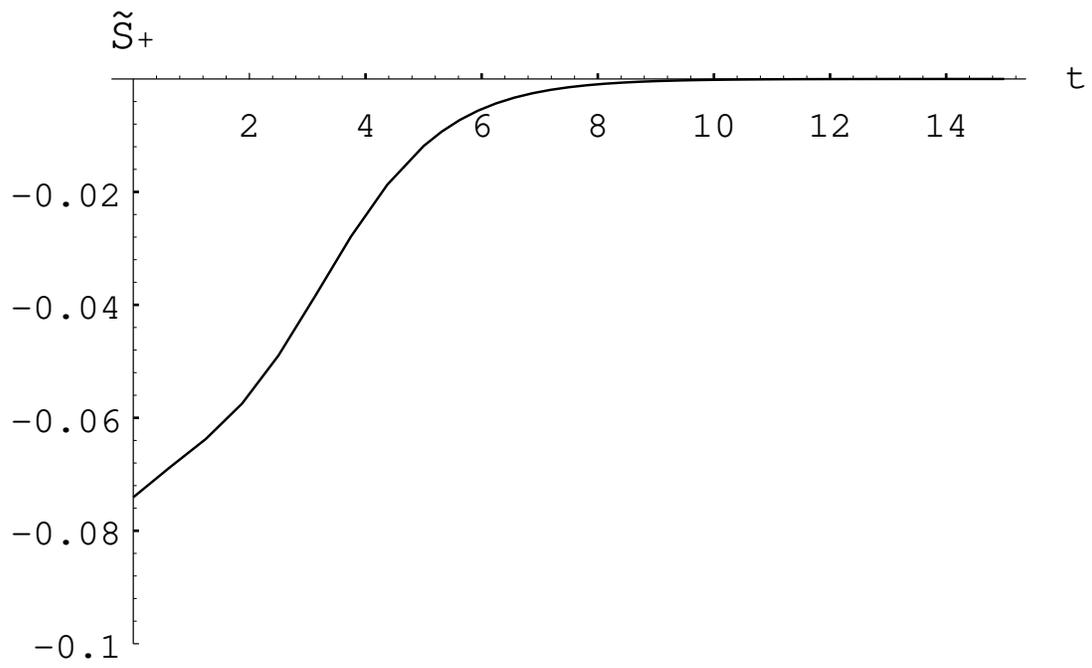
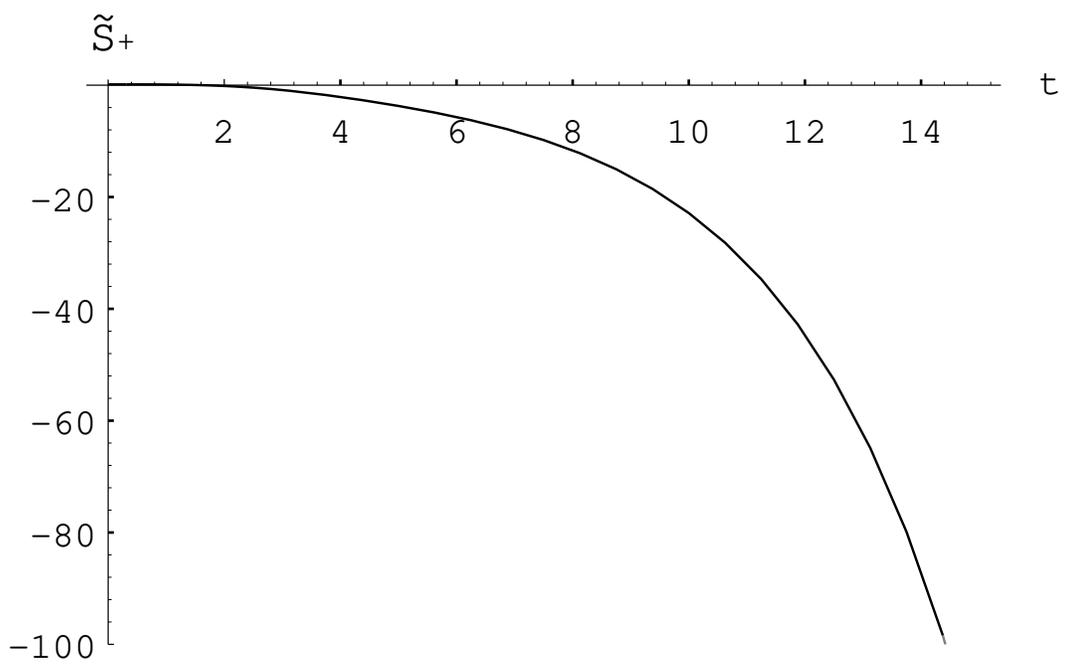
\end{document}